# Tractable fish growth models considering individual differences with an application to the fish *Plecoglossus altivelis altivelis*

Hidekazu Yoshioka[a, *], Yumi Yoshioka[b], Motoh Tsujimura[c]

[a] Japan Advanced Institute of Science and Technology, 1-1 Asahidai, Nomi, Ishikawa 923-1292, Japan

[b] Gifu University, Yanagido 1-1, Gifu, Gifu 501-1193, Japan.

[c] Doshisha University, Karasuma-Higashi-iru, Imadegawa-dori, Kamigyo-ku, Kyoto 602-8580, Japan

[*] Corresponding author: yoshih@jaist.ac.jp

ORCID:0000-0002-5293-3246 (Hidekazu Yoshioka), 0000-0002-0855-699X (Yumi Yoshioka), 0000-0001-6975-9304 (Motoh Tsujimura)

**Abstract**

Modeling fish growth is an important research topic in ecological and fishery sciences because body weight statistics directly affect the total biomass of fish in a habitat, which in turn affects their population dynamics. Many models of fish growth assume that the fish population in a habitat is homogenous, meaning that there is no physiological spectrum and, therefore, no size spectrum. Moreover, models that account for the size spectrum are not always analytically tractable. We present novel mathematical models of fish growth in which the body weight of each fish is assumed to follow a von Bertalanffy-type model whose proportionality coefficient, representing the maximum body weight, may differ among individual fish. This probabilistic description introduces the size spectrum into the model, owing to which the time-dependent probability density of this model is obtained explicitly. We also consider a misspecified version and a stochastic version of the model as advanced cases. We apply the first model to the real growth data of *Plecoglossus altivelis altivelis* as a keystone fish species in Japan. The model successfully reproduces the skewed size spectrum of this fish species over multiple years. We further use the stochastic model to investigate how fish growth dynamics are affected by environmental fluctuations.

***Disclosures and declarations***

**Acknowledgements:** The authors thank the Hii River Fisheries Cooperative for their valuable support in collecting body weight data for the individual fish used in this study. The authors could significantly improve the manuscript by the helpful comments from the reviewers.

**Funding:** This study was supported by the Japan Society for the Promotion of Science (KAKENHI, No. 22K14441), the Nippon Life Insurance Foundation (Environmental Problems Research Grant for Young Researchers, No. 24), and the Japan Science and Technology Agency (PRESTRO, No. JPMJPR24KE).

**Data availability statement:** The data will be made available upon reasonable request to the corresponding author.

**Competing interests:** The authors declare no competing interests relevant to the content of this article.

**Declaration of generative artificial intelligence (AI) in scientific writing:** The authors did not use generative AI technology to prepare this manuscript.

# 1. Introduction

## 1.1 Research background

Fishery resources are major natural resources for food and include ecologically important species in aquatic environments. Analyzing population dynamics, such as stock and growth dynamics, has been an important research topic in fisheries science and related research areas. Trout and salmon have been studied worldwide due to their high economic value. Individual-based models have been applied to the population dynamics of brown trout and Atlantic salmon in a river environment regulated by hydropower operations [1,2]. Interannual trends of Japanese anchovy abundance have been statistically analyzed to evaluate the sustainability of fish as a natural resource [3]. Fish passage impacts of paddlefish in a fragmented river environment have been investigated using a stochastic process model based on passage probabilities [4]. Additionally, the fish investigated in this study, *Plecoglossus altivelis altivelis* (*P. altivelis*), is a diadromous fish species that is commercially important and accounts for 6.7% of the total inland fish catch in Japan [5]. Both the stock and growth dynamics of *P. altivelis* have been studied in detail in mountainous river environments of Japan [6,7].

In this study, we focus on growth dynamics, specifically the temporal variation in the body weight in a water body because it is essential for estimating the biomass of fishery resources within it [8-10]. Growth curve models that generate sigmoidal shapes, such as logistic and von Bertalanffy (VB) models and their variants, have been proposed and investigated [11-13]. Growth curve models account for stress-response regulation [14], periodic environmental fluctuations [15], and stochastic fluctuations [16].

Growth curve models have been investigated from multiple angles. However, growth curve models that account for physiological heterogeneity among individuals in a population, which is naturally observed as the size spectrum (individual differences in the population), have not been well studied because of their higher complexity compared to homogenous ones. These models are often called size-structured models [17]. The size spectrum due to the coexistence between juvenile and adult fish has been pointed out to be essential for understanding the spatial distribution of certain marine fish species with an age structure [18]. The mortality rate of individual fish has been suggested to be size-dependent [19,20], and hence, the size spectrum is a crucial factor for assessing fish stocks. The existence of the size spectrum of a fishery resource implies uncertainty in catch during angling. In this view, it has been pointed out that high catch uncertainties would lead to higher fishing utility in recreational fishing [21].

Yoshioka [22] proposed the use of an uncertain logistic growth model in which the maximum body weight was assumed to differ among individuals, and the model was subsequently applied numerically to fisheries management problems involving *P. altivelis* [23]. However, the estimation procedure for the model parameters relies on a brute-force strategy that is inefficient and difficult to apply. An efficient, analytically tractable, and practical growth curve model that accounts for the size spectrum is necessary for operational applications in ecology and fisheries, such as integrating human behavior into ecological modeling [24]. More specifically, a size-structured model that can analytically reproduce the distribution of body weight or body length at any time would be efficient for use in fishery applications. Such a model would be more beneficial if it could be extended to an advanced version capable of addressing modeling errors and stochastic effects, which are often challenging in real problems. This motivation forms the basis of this study, which is detailed below.

## 1.2 Objective and contribution

The aim of this study is to propose and investigate an analytically tractable mathematical model for the growth dynamics of body weights of biological individuals in a physiologically heterogeneous population. This study also aims to apply the model and its extended versions to real data of the fish *P. altivelis* over multiple years. In our context, "size-structured" means that there are "individual differences" in the body weight. The contributions made to achieve these aims are the following.

Our model is based on the VB model for the growth of body weight of an individual fish in a homogeneous population [25]:

$$X_t = KL_t^3 \text{ with } L_t = 1-(1-u)e^{-rt}, \tag{1}$$

where $t \geq 0$ is time, $L_t$ is the (non-dimensional) body length of the fish at time $t$, $X_t$ is the body weight of the fish at time $t$, $r > 0$ is the growth rate of body length, $u \in (0,1)$ is a parameter characterizing the initial body weight through the relationship $X_0 = KL_0^3 = Ku^3$, and $K > 0$ is the maximum body weight. The power "3" in the first equation of (1) represents the well-established physiological scaling that the body weight is proportional to the body volume. The VB model (1) describes a sigmoidal growth when $u \in (0, 2/3)$, as proven in **Section 2**. VB-type models have been widely used in literature (e.g., [26-28]), whereas their size-structured version has not been extensively investigated, to the best of our knowledge.

The main contribution of this study is the proposal of a body-weight model for individual fish in a heterogeneous population. The size-structured version proposed in this study is as follows:

$$X_t^{(i)} = K^{(i)} L_t^3, \tag{2}$$

where the superscript $(i)$ represents the $i$ th individual in the population. We assume that the elements of the sequence $K^{(i)}$ are mutually independent and follows a common probability density function (PDF) $p = p(K)$ ($K > 0$). In the present model, the heterogeneity of body weight is assumed to be encoded by the PDF. The advantage of this model is its analytical tractability, such that the PDF of the body weight of individuals in the population is expressed using $p$ and $L_t$ through relationship (2). This characteristic of the model is exploited in the identification procedure, where we assume a Gamma or inverse Gamma distribution for $p$, with which the PDF of the body weight is obtained analytically. This is a significant advantage of the proposed model, which supports its usability in various applications.

Another advantage of the proposed model is that a misspecified or stochastic version can be constructed efficiently. Here, misspecification means that the PDF of the maximum body weight $p$ is distorted by modeling and/or measurement errors, which eventually affects the statistical analysis of the proposed model. We propose evaluating the misspecification through the relative entropy known as the Kullback–Leibler divergence as well [29,30], which is a widely used measure between probabilistic models. We show that the statistics of the distorted model that give the worst-case upper or lower bound of the mean body weight are obtained analytically. We also show that the modeler's uncertainty aversion critically affects the distorted model, possibly causing the worst-case upper bound to diverge, making the distorted model meaningless. A discussion of the Orlicz space [31,32] rigorously provides a sharp condition to prevent this peculiar phenomenon from occurring.

The stochastic fluctuation here means that each $L_t$ for each individual fish is governed by a stochastic differential equation (SDE), where the noise conceptually represents environmental fluctuations, such as the hydrodynamic disturbance in the water body where the fish live. We assume that $L_t$ follows a Jacobi process, resulting in the simplest well-defined SDE whose solution is bounded between 0 and 1 (Chapter 6 in [33]), consistent with the deterministic case. The Jacobi process allows us to obtain statistical moments of body weight by solving a system of linear ordinary differential equations (ODEs) efficiently. Therefore, we do not need any statistical simulations, even for the stochastic version. In the literature, SDEs have been applied to biological phenomena, including individuals, using continuous [34-38] and jump-driving noise [39]. In contrast, models that consider both physiological differences (static uncertainty) and stochastic fluctuations (dynamic uncertainty) are scarce. We show that a framework considering each type of randomness inherits the analytical tractability of the proposed model.

Finally, we apply the proposed model to the collected body weight data of the fish *P. altivelis* in the Hii River, Japan. Fish have a one-year life cycle, and the assumption of not having an age structure is satisfied (e.g., [6, 40, 41]). We only have weight data but not length data, so we regard the variable $L$ as an auxiliary one to explain the temporal variation of the PDF of body weight. The data collected suggest that the size spectrum, namely, the PDF of body weight, is unimodal and positively skewed, both of which are successfully captured by the proposed model. The influences of rising water temperatures on the biological growth of fish are also discussed using collected and public temperature data to potentially advance temperature-dependent fish growth modeling (e.g., [42-44]). We also consider hypothetical scenarios to evaluate the influences of PDF misspecifications and environmental fluctuations. We then show that the misspecification would have influences while the stochastic fluctuation would not, at least in our case study, because the stochastic fluctuation has been identified as being small.

### 1.3 Structure of this paper

The rest of this paper is organized as follows: **Section 2** analyzes the VB and proposed models in more detail. Models subject to misspecifications and stochastic fluctuations are also introduced in this section. The key differences between existing and proposed models are discussed in this section as well. **Section 3** presents the data collected of *P. altivelis*. **Section 4** presents the identification and investigation of the proposed model and further numerical studies of advanced models. **Section 5** concludes the paper and discusses future perspectives. **Appendices** present supplementary results.

## 2. Fish growth model

### 2.1 VB model

The classical VB model governs the body length $l_t$ of a biological individual through a linear ODE (e.g., [25])

$$\frac{\mathrm{d}l_t}{\mathrm{d}t} = r\left(l_\infty - l_t\right) \text{ or equivalently } l_t = l_\infty - \left(l_\infty - l_0\right)e^{-rt}, \quad t > 0, \tag{3}$$

where $l_0 > 0$ is the length at the initial time $t = 0$, $l_\infty \left(> l_0\right)$ is the asymptotic length that is real and finite. We normalize $l_t$ as $L_t = l_t / l_\infty$ with $u = l_0 / l_\infty \in \left(0,1\right)$ and obtain the second equation of (1). Then we have $0 \le L_t \le 1$ at any time $t \ge 0$. The bodyweight $X_t$ of an individual at time $t$ is then defined as in the first equation of (1) by assuming a common cubic allometric relationship (e.g., [6]).

The VB model can capture sigmoidal growth. Indeed, an elementary calculation yields

$$\frac{1}{K}\frac{\mathrm{d}X_t}{\mathrm{d}t} = \frac{\mathrm{d}\left(L_t\right)^3}{\mathrm{d}t} = 3r\left(L_t\right)^2\left(1 - L_t\right) \ge 0 \tag{4}$$

and

$$\begin{aligned}\frac{1}{K}\frac{\mathrm{d}^2 X_t}{\mathrm{d}t^2} &= 3r\frac{\mathrm{d}}{\mathrm{d}t}\left(\left(L_t\right)^2 - \left(L_t\right)^3\right) \\ &= 3r\left(2 - 3L_t\right)L_t\frac{\mathrm{d}L_t}{\mathrm{d}t} \\ &= 3r^2\left(1 - L_t\right)L_t\left(2 - 3L_t\right) \\ &= 3r^2\left(1 - L_t\right)L_t\left(2 - 3\left(1 - \left(1 - u\right)e^{-rt}\right)\right) \\ &= 3r^2\left(1 - L_t\right)L_t e^{-rt}\left(3\left(1 - u\right) - e^{rt}\right)\end{aligned}. \tag{5}$$

We have $3\left(1-u\right) > 1$ if and only if $u \in \left(0, 2/3\right)$. Under this assumption, according to (5), there is an inflection point of $X_t$ at the positive time $t = \frac{1}{r}\ln\left(3\left(1-u\right)\right)$, and hence $X_t$ has a sigmodal shape.

### 2.2 Proposed VB model with size spectrum

We consider a randomized version of the VB model that accounts for the size spectrum of individuals in a population as

$$X_t^{(i)} = K^{(i)}L_t^3, \quad i \in \left\{1, 2, 3, ..., I\right\}. \tag{6}$$

Here, the superscript $\left(i\right)$ represents the $i$ th individual in the population, where we assume that there exists a large number of individuals ($I >> 1$, with $I$ being the total number of individuals). We assume that each $K^{(i)}$ is mutually independent and follows a common PDF $p = p\left(K\right)$ ($K > 0$). All individuals are assumed to have the common deterministic factor $L_t^3$, and hence the size spectrum is encoded in $p$. Mathematically, the variable $L_t$ needs not to be the body length but can be an auxiliary variable to reproduce possibly sigmoidal growth dynamics, which is our position in this paper.

Given a PDF $p$ of the maximum body weight, the body weight $X_t$ of a representative individual at time $t$ satisfies

$$\Pr\left(X_t \le x\right) = \Pr\left(KL_t^3 \le x\right) = \Pr\left(K \le \frac{x}{L_t^3}\right),\quad x > 0. \tag{7}$$

The PDF $q_t = q_t\left(x\right)$ of $X_t$ is then derived as follows if $p$ is known:

$$q_t\left(x\right) = \frac{\mathrm{d}}{\mathrm{d}x}\Pr\left(X_t \le x\right) = \frac{1}{L_t^3} p\left(\frac{x}{L_t^3}\right),\quad x > 0. \tag{8}$$

Therefore, the PDF $q_t$ is obtained explicitly if $p$ is. We assume Gamma or inverse Gamma distribution for PDF $p$, whose statistics can be computed analytically. In addition, the observation results suggest that the PDFs of the body weight are unimodal and have longer right tails (positive skewness): for $x > 0$,

**(Gamma distribution)**
$$p\left(x\right) = \frac{1}{\Gamma\left(\alpha\right)}\frac{x^{\alpha-1}}{\beta^{\alpha}}\exp\left(-\frac{x}{\beta}\right) \tag{9}$$

and

**(Inverse Gamma distribution)**
$$p\left(x\right) = \frac{1}{\Gamma\left(\alpha\right)}\frac{\beta^{\alpha}}{x^{\alpha+1}}\exp\left(-\frac{\beta}{x}\right). \tag{10}$$

Here, $\Gamma\left(\alpha\right) = \int_0^{+\infty} x^{\alpha-1}e^{-x}\mathrm{d}x$ is the Gamma function, $\alpha > 0$ is the shape parameter, and $\beta > 0$ is the scale parameter.

Average, standard deviation, coefficient of variation, and skewness of the two distributions are obtained using elementary calculations and are listed in **Table 1**. For the inverse Gamma distribution, higher-order statistics exist only for sufficiently high values of the shape parameter $\alpha$. Indeed, a crucial difference between the two distributions is that the tail of the PDF decays almost exponentially in the Gamma distribution, whereas it decays polynomially in the inverse Gamma one; the latter therefore has a heavier tail. This qualitative difference plays a vital role in an extended growth curve model in which the PDF $p$ is assumed to be misspecified, as explained in the next subsection.

**Table 1.** Average, standard deviation, coefficient of variation, and skewness of the Gamma and inverse Gamma distributions.

| Statistics | Gamma distribution | Inverse Gamma distribution |
|---|---|---|
| Average | $\alpha\beta$ | $\frac{\beta}{\alpha-1}$ |
| Standard deviation | $\sqrt{\alpha}\beta$ | $\frac{\beta}{(\alpha-1)\sqrt{\alpha-2}}$ if $\alpha > 2$ |
| Coefficient of variation | $\frac{1}{\sqrt{\alpha}}$ | $\frac{1}{\sqrt{\alpha-2}}$ if $\alpha > 2$ |
| Skewness | $\frac{2}{\sqrt{\alpha}}$ | $\frac{4\sqrt{\alpha-2}}{\alpha-3}$ if $\alpha > 3$ |

### 2.3 Model subject to misspecification

#### 2.3.1 Relative entropy

The parameters in a biological model must be typically identified from a limited amount of data, which may contain measurement errors. Therefore, an identified model may deviate from the true model if it exists. We present a modeling framework to evaluate the statistics of the proposed model, especially the average directly affecting the population biomass. In this subsection, we assume $p(K) > 0$ for any $K > 0$, which is at least satisfied in our application study in **Section 4**.

We consider a situation in which the difference between the true PDF $\hat{p} = \hat{p}(K)$, assuming that it exists, and the estimated PDF $p = p(K)$ from some data, can be evaluated through the relative entropy $\mathbb{D}(p, \hat{p})$ [29]:

$$\mathbb{D}(p, \hat{p}) = \int_0^{+\infty} \left(\phi(x)\ln\phi(x) - \phi(x) + 1\right) p(x)\,\mathrm{d}x , \tag{11}$$

where $\phi = \hat{p} / p > 0$. The function $y \ln y - y + 1$ ($y \geq 0$) is non-negative, convex, and admits a global minimum value of 0 at $y = 1$, where we set $0 \ln 0 = 0$. This implies that the relative entropy is nonnegative and vanishes if and only if $p = \hat{p}$, that is, if the estimated model is identical to the true model. Relative entropy is an asymmetric measure used to evaluate the difference between two PDFs because of $\mathbb{D}(p, \hat{p}) \neq \mathbb{D}(\hat{p}, p)$ in general. The use of $\mathbb{D}(p, \hat{p})$ but not $\mathbb{D}(\hat{p}, p)$ is justified in our approach that $p$ is the identified model from data and $\hat{p}$ is the alternative one, and the difference is measured with respect to $p$.

#### 2.3.2 Worst-case upper-bound

An advantage of using relative entropy to evaluate possible misspecifications of the proposed model is that the worst-case overestimation and underestimation of the average body weight can be obtained analytically if we assume a Gamma-type $p$. The worst-case upper-bound $\bar{K}(\varepsilon)$ of the average maximum body weight given a relative entropy bound is defined as follows: for any bound $\varepsilon > 0$,

$$\bar{K}(\varepsilon) = \max_{\hat{p}} \left\{ \underbrace{\int_0^{+\infty} x\hat{p}(x)\,\mathrm{d}x}_{\text{Average evaluated by an alternative model}} \middle| \underbrace{\hat{p}(\cdot) \geq 0,\ \int_0^{+\infty} \hat{p}(x)\,\mathrm{d}x = 1,}_{\text{Maximizer should be a PDF}} \ \underbrace{\mathbb{D}(p, \hat{p}) \leq \varepsilon}_{\text{Bound of relative entropy}} \right\}. \tag{12}$$

This type of formulation has been frequently employed in the literature on decision theory and related research areas (e.g., [45-49]) and is applied to the proposed model owing to its generality. Using the classical Lagrangian multiplier technique, this maximization problem can be recast as an optimization problem without the constraint of relative entropy:

$$\bar{K}(\varepsilon) = \max_{\hat{p}} \left\{ \int_0^{+\infty} x\hat{p}(x)\,\mathrm{d}x - \eta\mathbb{D}(p, \hat{p}) \middle| \hat{p}(\cdot) \geq 0,\ \int_0^{+\infty} \hat{p}(x)\,\mathrm{d}x = 1 \right\}. \tag{13}$$

Here, $\eta = \eta(\varepsilon) > 0$ is a function of $\varepsilon$, and is the Lagrangian multiplier corresponding to the sensitivity of the modeler to model misspecification; a smaller value of $\eta$ implies less aversion of the modeler against misspecification in the PDF $p$. The maximizer $p^*$ of (13) is obtained by a straightforward calculation as follows (e.g., Section 4.1 in [45]):

$$p^*(x) = \frac{p(x)e^{\frac{x}{\eta}}}{\int_0^{+\infty} p(x)e^{\frac{x}{\eta}}\mathrm{d}x}, \quad x > 0. \tag{14}$$

The denominator exists if $\eta > \beta$ as for (16) presented later. Then, we have

$$\bar{K}(\varepsilon) = \eta \ln\left(\int_0^{+\infty} p(x)e^{\frac{x}{\eta}}\mathrm{d}x\right) \tag{15}$$

if the denominator of (14) exists. We need to find the functional form of $\eta = \eta(\varepsilon)$, which is possible if we assume a Gamma-type $p$ (for the inverse Gamma-type $p$, this is actually impossible because of its theoretically limitation explained in **Remark 2** presented later). In this case, if $\eta > \beta$, a straightforward calculation shows that

$$\bar{K}(\varepsilon) = \eta\alpha \ln\left(\frac{1}{1-\beta/\eta}\right) \text{ and } \int_0^{+\infty} xp^*(x)\mathrm{d}x = \eta\alpha\left(\frac{1}{1-\beta/\eta} - 1\right), \tag{16}$$

and hence

$$\mathbb{D}(p, p^*) = \frac{1}{\eta}\left(\int_0^{+\infty} xp^*(x)\mathrm{d}x - \bar{K}(\varepsilon)\right) = \alpha\left(\frac{1}{1-\beta/\eta} - 1 - \ln\left(\frac{1}{1-\beta/\eta}\right)\right). \tag{17}$$

Both equations in (16) are obtained by directly substituting (14) into (15) (the left equation) and the definition of the average (the right one) along with the fact that $p$ is the PDF of the Gamma distribution. According to (17), $\mathbb{D}(p, p^*)$ as a function of $\eta \in (\beta, +\infty)$ strictly decreases from $+\infty$ to 0. Therefore, by the classical intermediate value theorem, for each $\varepsilon > 0$, there exists a unique solution $\eta = \bar{\eta} > 0$ to the equation

$$\alpha\left(\frac{1}{1-\beta/\eta} - 1 - \ln\left(\frac{1}{1-\beta/\eta}\right)\right) = \varepsilon, \tag{18}$$

which implicitly defines the desired relationship $\bar{\eta} = \bar{\eta}(\varepsilon)$. We have $\lim_{\varepsilon \to +0} \bar{K}(\varepsilon) = \lim_{\bar{\eta} \to +\infty} \bar{K}(\varepsilon) = \alpha\beta$.

In the present framework, the worst-case relative entropy (17) is proportional to the shape parameter $\alpha$ and depends on $\beta, \eta$ only through the ratio $\beta/\eta$. Proportionality with respect to $\alpha$ suggests that the regularity, and hence the functional form, of the Gamma distribution controls the possible misspecification. A larger $\alpha$ suggests the existence of the global maximum that is more distant from the origin. From an engineering viewpoint, this parameter dependence suggests that investigations of the populations having a dominant mode in the body weight distribution need more effort. The same reasoning applies to the lower-bound explained later.

Given $\varepsilon > 0$ and hence $\bar{\eta} = \bar{\eta}(\varepsilon) > \beta$, the worst-case upper bound of the average body weight $\bar{X}_t(\varepsilon)$ at time $t \geq 0$ is obtained as

$$\bar{X}_t(\varepsilon) = \bar{K}(\varepsilon) L_t^3. \tag{19}$$

Following the analytical procedure described in **Section 2.3**, the PDF $\bar{q}_t$ that represents a misspecified model to give the upper-bound $\bar{K}(\varepsilon)$ is obtained as follows due to (14):

$$\bar{q}_t(x) = C \frac{1}{L_t^3} p\left(\frac{x}{L_t^3}\right) \exp\left(\frac{x}{\bar{\eta} L_t^3}\right), \quad x > 0 \tag{20}$$

with $C > 0$ being the normalization constant to enforce $\int_0^{+\infty} \bar{q}_t(x) \mathrm{d}x = 1$. The equation (20) has been obtained by simply considering (8) where $p$ is replaced by $p^*$. Consequently, the proposed model subject to misspecification can be fully dealt with analytically.

Finally, we obtain the following **Proposition 1**:

***Proposition 1***

*The worst-case upper bound $\bar{K}$ is obtained as (16), and the PDF of the body weight that archives $\bar{K}$ is given by (20). Moreover, (18) implicitly defines the relation $\bar{\eta} = \bar{\eta}(\varepsilon)$.*

#### 2.3.3 Worst-case lower-bound

Symmetrically, the worst-case lower-bound $\underline{K}(\varepsilon)$ of the average maximum body weight given a relative entropy bound is defined as follows: for each $\varepsilon > 0$,

$$\underline{K}(\varepsilon) = \min_{\hat{p}} \left\{ \underbrace{\int_0^{+\infty} x \hat{p}(x) \mathrm{d}x}_{\text{Average evaluated by an alternative model}} \middle| \underbrace{\hat{p}(\cdot) \geq 0,\ \int_0^{+\infty} \hat{p}(x) \mathrm{d}x = 1,}_{\text{Minimizer should be a PDF}} \ \underbrace{\mathbb{D}(p, \hat{p}) \leq \varepsilon}_{\text{Bound of relative entropy}} \right\} \tag{21}$$

along with the multiplier form

$$\underline{K}(\varepsilon) = \min_{\hat{p}} \left\{ \int_0^{+\infty} x \hat{p}(x) \mathrm{d}x + \eta \mathbb{D}(p, \hat{p}) \middle| \hat{p}(\cdot) \geq 0,\ \int_0^{+\infty} \hat{p}(x) \mathrm{d}x = 1 \right\}. \tag{22}$$

The minimizer is given by (the minimization problem is solved similarly to the maximization one because a minimization problem of a function can be written a maximization problem of a minus of the same function, e.g., Section 4.1 in [45])

$$p^*(x) = \frac{p(x) e^{\frac{-x}{\eta}}}{\int_0^{+\infty} p(x) e^{\frac{-x}{\eta}} \mathrm{d}x}, \quad x > 0 \tag{23}$$

and we have

$$\underline{K}(\varepsilon) = -\eta \ln\left( \int_0^{+\infty} p(x) e^{\frac{-x}{\eta}} \mathrm{d}x \right). \tag{24}$$

Note that the denominator of (23) always exists. For any $\eta = \eta(\varepsilon) > 0$, we obtain

$$\underline{K}(\varepsilon) = -\eta\alpha \ln\left(\frac{1}{1+\beta/\eta}\right) \text{ and } \int_0^{+\infty} xp^*(x)\,\mathrm{d}x = \eta\alpha\left(1 - \frac{1}{1+\beta/\eta}\right), \tag{25}$$

and hence

$$\mathbb{D}(p, p^*) = \frac{1}{\eta}\left(\underline{K}(\varepsilon) - \int_0^{+\infty} xp^*(x)\,\mathrm{d}x\right) = \alpha\left(-\ln\left(\frac{1}{1+\beta/\eta}\right) + \frac{1}{1+\beta/\eta} - 1\right). \tag{26}$$

The right-most side of (26) is strictly decreasing with respect to $\eta > 0$ and has the limit 0 for $\eta \to +\infty$ and the limit $+\infty$ for $\eta \to +0$. Therefore, by the classical intermediate value theorem, for each $\varepsilon > 0$ there exists a unique solution $\underline{\eta} > 0$ to the equation

$$\alpha\left(\frac{1}{1+\beta/\eta} - 1 - \ln\left(\frac{1}{1+\beta/\eta}\right)\right) = \varepsilon, \tag{27}$$

which implicitly defines the desired relationship $\underline{\eta} = \underline{\eta}(\varepsilon)$.

Given $\varepsilon > 0$ and hence $\underline{\eta} = \underline{\eta}(\varepsilon) > 0$, the worst-case lower bound of the average body weight $\underline{X}_t(\varepsilon)$ at time $t \geq 0$ is obtained as

$$\underline{X}_t(\varepsilon) = \underline{K}(\varepsilon) L_t^3. \tag{28}$$

Moreover, the PDF $\underline{q}_t$ corresponding to the misspecified model, is derived as

$$\underline{q}_t(x) = C\frac{1}{L_t^3} p\left(\frac{x}{L_t^3}\right)\exp\left(\frac{-x}{\underline{\eta} L_t^3}\right), \quad x > 0 \tag{29}$$

with $C > 0$ being a normalization constant. Finally, **Table 2** shows that the prediction under the worst-case misspecification yields smaller average and variance, while the coefficient of variation and skewness are not affected as in the upper-bounding case.

Consequently, we obtain the following **Proposition 2**:

***Proposition 2***

*The worst-case lower bound* $\underline{K}$ *is obtained as (25), and the PDF of the body weight that archives* $\underline{K}$ *is given by (29). Moreover, (27) implicitly defines the relation* $\underline{\eta} = \underline{\eta}(\varepsilon)$.

Finally, the statistics of the distorted Gamma PDF are explicitly obtained, as summarized in **Table 2**. A comparison between **Tables 1-2** shows that the prediction under the worst-case misspecification yields a larger average and variance for the upper-bound and a smaller average and variance for the lower-bound, whereas the coefficients of variation and skewness are not affected. This is due to the distorted PDFs do not affect the polynomial term and, hence, the regularity of the original Gamma PDF. Remarkably, the upper-bound (respectively, the lower bound) of the average and standard deviation are proportional to the common factor $\frac{1}{1-\beta/\eta}$ (respectively, $\frac{1}{1+\beta/\eta}$). More specifically, the sensitivity of these statistics is solely controlled by the $\beta$ for each $\eta$, suggesting that a model with a larger $\beta$ is more susceptible to misspecification.

**Table 2.** Average, standard deviation, coefficient of variation, and skewness of the distorted Gamma distribution.

| Statistics | Upper-bounding case | Lower-bounding case |
|---|---|---|
| Average | $\alpha\beta\times\frac{1}{1-\beta/\eta}$ | $\alpha\beta\times\frac{1}{1+\beta/\eta}$ |
| Standard deviation | $\sqrt{\alpha}\beta\times\frac{1}{1-\beta/\eta}$ | $\sqrt{\alpha}\beta\times\frac{1}{1+\beta/\eta}$ |
| Coefficient of variation | $\frac{1}{\sqrt{\alpha}}$ | $\frac{1}{\sqrt{\alpha}}$ |
| Skewness | $\frac{2}{\sqrt{\alpha}}$ | $\frac{2}{\sqrt{\alpha}}$ |

***Remark 1***

As demonstrated above, the proposed modeling framework of the size-structured VB type, subject to misspecification, offers an analytically tractable methodology for statistically evaluating the body weight dynamics of individual fish. However, one should consider that the worst-case upper bound becomes meaningless; $\bar{K}(\varepsilon)=+\infty$ if $\eta\leq\beta$, that is, if the modeler sets a sufficiently high aversion to the misspecification of the PDF $p$. Mathematically, this is a consequence of

$$\int_0^{+\infty} e^{\frac{x}{\eta}} p(x)\mathrm{d}x \propto \int_0^{+\infty} x^{\alpha-1} e^{\frac{x}{\eta}-\frac{x}{\beta}}\mathrm{d}x = +\infty \tag{30}$$

for $\eta\leq\beta$. For the lower bound, such a technical issue does not arise because of

$$\int_0^{+\infty} e^{-\frac{x}{\eta}} p(x)\mathrm{d}x \propto \int_0^{+\infty} x^{\alpha-1} e^{-\frac{x}{\eta}-\frac{x}{\beta}}\mathrm{d}x < +\infty . \tag{31}$$

This is why we need to restrict the range of the Lagrangian multiplier $\eta$ to $(\beta,+\infty)$ for the upper-bound.

***Remark 2***

The reason that we did not apply the inverse Gamma distribution is due to the theoretical limitation that the calculations of the worst-case upper-bound are meaningless, i.e., we always obtain $\bar{K}=+\infty$. More specifically, we did not deal with the inverse Gamma-type $p$ in the previous subsection because of

$$\int_0^{+\infty} e^{\frac{x}{\eta}} p(x)\mathrm{d}x \propto \int_0^{+\infty} \frac{1}{x^{\alpha+1}} e^{\frac{x}{\eta}-\frac{\beta}{x}}\mathrm{d}x = +\infty \tag{32}$$

for any $\eta > 0$. The worst-case upper bound is therefore always $+\infty$, and hence, the problem itself is meaningless; even a small model misspecification becomes critical for the inverse Gamma case.

***Remark 3***

Finally, the divergence phenomenon (32) from a mathematical viewpoint implies that the function $x$ does not belong to the Orlicz space associated with the inverse Gamma-type PDF and exponential Orlicz function $\exp(\cdot)$ (e.g., Part IV in [31]. See also **Section A.4** in **Appendix**), i.e., the divergence occurs for any $\eta > 0$. In contrast, the divergence phenomenon (30) implies that the function $x$ belongs to the Orlicz space associated with the Gamma-type PDF and exponential Orlicz function $\exp(\cdot)$, i.e., the divergence does not occur for any $\eta > 0$. In particular, the tail behavior of the PDEs plays a role; Gamma and inverse Gamma PDFs exhibit exponential and polynomial decays, respectively. This subtle difference may not be important from an engineering standpoint but can be crucial when dealing with a similar but more complex model whose explicit treatment is not easy, such that sophisticated mathematical analysis becomes necessary.

***Remark 4***

A mathematical model that is more realistic than those studied in this study should incorporate some evolution equation (a governing equation of the fish in multiple years) where a growth model should be incorporated into it. In this view, the lack of the evolution equation is a limitation of this study. Addressing this issue will need some additional field investigations of the population dynamics including reproduction and migration of the fish, which is currently undergoing by our study.

***Remark 5***

As discussed above, considering the worst-case upper-bound (resp., lower-bound) in the sense of the relative entropy leads to a model that most overestimates (resp., underestimates) the growth of individuals within the prescribed size of the relative entropy. Because the relative entropy corresponds to some modeling error, they lead to the most overestimating and underestimating models provided that the degree of modeling errors is given. Moreover, the use of the relative entropy is advantageous for our model because the alternative model corresponding to the worst-case again admits a Gamma-type PDF (with different parameter values). This implies that the analytical results applicable to the model without misspecifications remain usable. By contrast, a difficulty in application, as pointed out in the comment, is that the true model is usually unknown, and hence it is difficult to estimate the proper relative entropy value. This is the reason we have conducted some sensitivity analysis about the errors in $\beta$ (**Section 4**), to address the aforementioned limitation.

### 2.4 Model with stochastic fluctuation

The model explained above does not account for environmental fluctuations, such as temporal variations in flow discharge, water depth, water temperature, and local hydrodynamic disturbances. We propose a stochastic version of the model presented in **Section 2.2**. This model is based on (6) but with the process $L_t^{(i)}$ for each individual governed by the SDE defined in the Itô's sense

$$\mathrm{d}L_t^{(i)} = r\left(1-L_t^{(i)}\right)\mathrm{d}t + \sigma\sqrt{L_t^{(i)}\left(1-L_t^{(i)}\right)}\mathrm{d}B_t^{(i)},\quad t>0 \tag{33}$$

with the additional parameter $\sigma > 0$ representing the stochasticity and $B^{(i)} = \left(B_t^{(i)}\right)_{t\geq 0}$ each (independent from the randomness generating $K$ ) is a mutually-independent standard 1-D Brownian motion as a canonical white noise process. Equation (33) is a Jacobi process as the simplest noise model whose solution is bounded between 0 and 1 if $\sigma^2 \leq 2r$ (Chapter 6 in [33]). The rationale behind the SDE (33) is that we can reproduce a stochastically fluctuating $L_t$ when $\sigma \neq 0$ and this model reduces to the original one when there is no noise $\sigma = 0$. Moreover, we assume that the noise, the Brownian motion, is mutually independent among all individuals in the population by assuming that all individuals face different fluctuation scenarios.

The SDE (33) is therefore a plausible stochastic model in the sense that the normalized body length $L_t^{(i)}$ can be rigorously kept between 0 and 1 as in the deterministic version presented earlier. This is due to the specific form of the diffusion coefficient multiplied by the Brownian motion term. Moreover, the motivation of using the specific form of the diffusion coefficient from an application viewpoint is that the moments of $L_t^{(i)}$ can be found by solving a system of liner ODEs, and even exactly for low-order moments as shown in **Section A.1** in **Appendix**. This property will be lost if we replace $\sqrt{L_t^{(i)}\left(1-L_t^{(i)}\right)}$ by seemingly simpler ones such as $L_t^{(i)}\left(1-L_t^{(i)}\right)$. Thus, the form of the SDE (33) can be justified both from theoretical and practical viewpoints. By contrast, a drawback of the SDE (33) is that investigating its path-wise properties need some Monte-Carlo simulation that may be computationally extensive. Nevertheless, this drawback does not apply to this study because our fitting method presented in **Section 4** is based mainly on the moments. If one is interested in a possibly more realistic stochastic model, then it will be more complex than ours, and hence its treatment will be more difficult both from theoretical and computational standpoints.

### 2.5 Key differences from existing approaches

The key differences between the conventional and proposed models are summarized as follows. First, there exist few models that account for the size structure in a VB model, where the length and growth rate ( $u$ and $r$ in our context) [50] and environmental limitation effect [51] are randomized (the first "1" in the right-hand side of the second equation of (1) is replaced by a random variable). By contrast, randomness is assumed in modeling the maximum body weight in this study, with which a fully analytically tractable model was obtained. Some growth curves with uncertain parameter values are based on a logistic model, in which the maximum body weight is assumed to be distributed among individuals in a population [52,53].

To the best of our knowledge, such models have been unified within the class of random differential equations [54], whereas the model of our type has not been investigated. A disadvantage of the VB-type model compared to the logistic model is that the latter can cover the entire life history of an individual fish, while the former does not. A logistic model in our context can be expressed as

$$X_t = \frac{K}{1+\left(K/X_0 - 1\right)e^{-rt}}, \tag{34}$$

which is always positive for any $t \in (-\infty, +\infty)$. By contrast, the classical VB model becomes meaningless at time $L_t = 0$, namely, at time $\frac{1}{r}\ln(1-u)$. Therefore, the VB model is defined only for $t \in \left(\frac{1}{r}\ln(1-u), +\infty\right)$. Because the initial time 0 is determined by the modeler, this technical issue would be critical if the modeler wanted to investigate the entire life history of biological individuals. In this study, we apply the proposed models to *P. altivelis* from summer to autumn during which the fish live in a river. This period does not include the larval period of the fish.

Concerning the evaluation of misspecification, theoretically, the optimization-based method to evaluate the upper and lower bounds of the average body weight proposed in the previous subsection can be applied to the models reviewed above. However, the applicability of this method should be carefully considered because the upper bound may be meaningless ($\bar{K}(\varepsilon) = +\infty$) (e.g., for the inverse Gamma-type $p$). This ill-posedness issue can be partly mitigated by using a suitable genderized relative entropy, called *f*-divergence or *φ*-divergence (e.g., [55,56]), as theoretically suggested in the literature [32]. A drawback of this remedy is the increase of model parameters in the relative entropy and the loss of analytical tractability. We do not further address this issue in this study.

## 3. Real data

### 3.1 Studied river and target species

The target fish species in this study is the diadromous migratory fish *P. altivelis* in the Hii River system, which is a Class A river system in Japan (**Figure 1**). The main branch of the Hii River is 153 km long, with a watershed area of 2,540 km$^2$. It originates from Mt. Sentsu at an altitude of 1,143 m in Okuizumo-cho, Shimane Prefecture, flows through Lake Shinji, Lake Nakaumi, and empties into the Sea of Japan[1]. The Mitoya River is the largest tributary of the Hii River.

The diadromous fish *P. altivelis* is a major inland fishery resource in the Hii River system. Most of the fish analyzed in this study were captured in the Hii and Mitoya Rivers. The common lifespan of the fish is one year, and they migrate between the sea and rivers. They spend their larval and early juvenile stages (November-March) in the sea and then migrate upstream to nearby rivers in spring (April to May). They grow in the midstream sections of rivers from summer to autumn (June to October). In autumn (October to November), they descend to the downstream areas of the rivers to spawn and die (e.g., [6,40,41]). The full life history of the fish is still unknown owing to the lack of data during the larval and early juvenile stages in winter. They are thought to overwinter in Lake Shinji, Lake Nakaima, or the Sea of Japan. The fishing season for *P. altivelis* in the Hii River system runs from July 1 to December 31 each year. Water and air temperatures were measured in and around the Hii River, as shown in **Figure 1**.

[1] Ministry of Land, Infrastructure, Transport and Tourism
https://www.mlit.go.jp/river/toukei_chousa/kasen/jiten/nihon_kawa/0713_hiikawa/0713_hiikawa_00.html
(in Japanese. Last accessed on November 27, 2024)

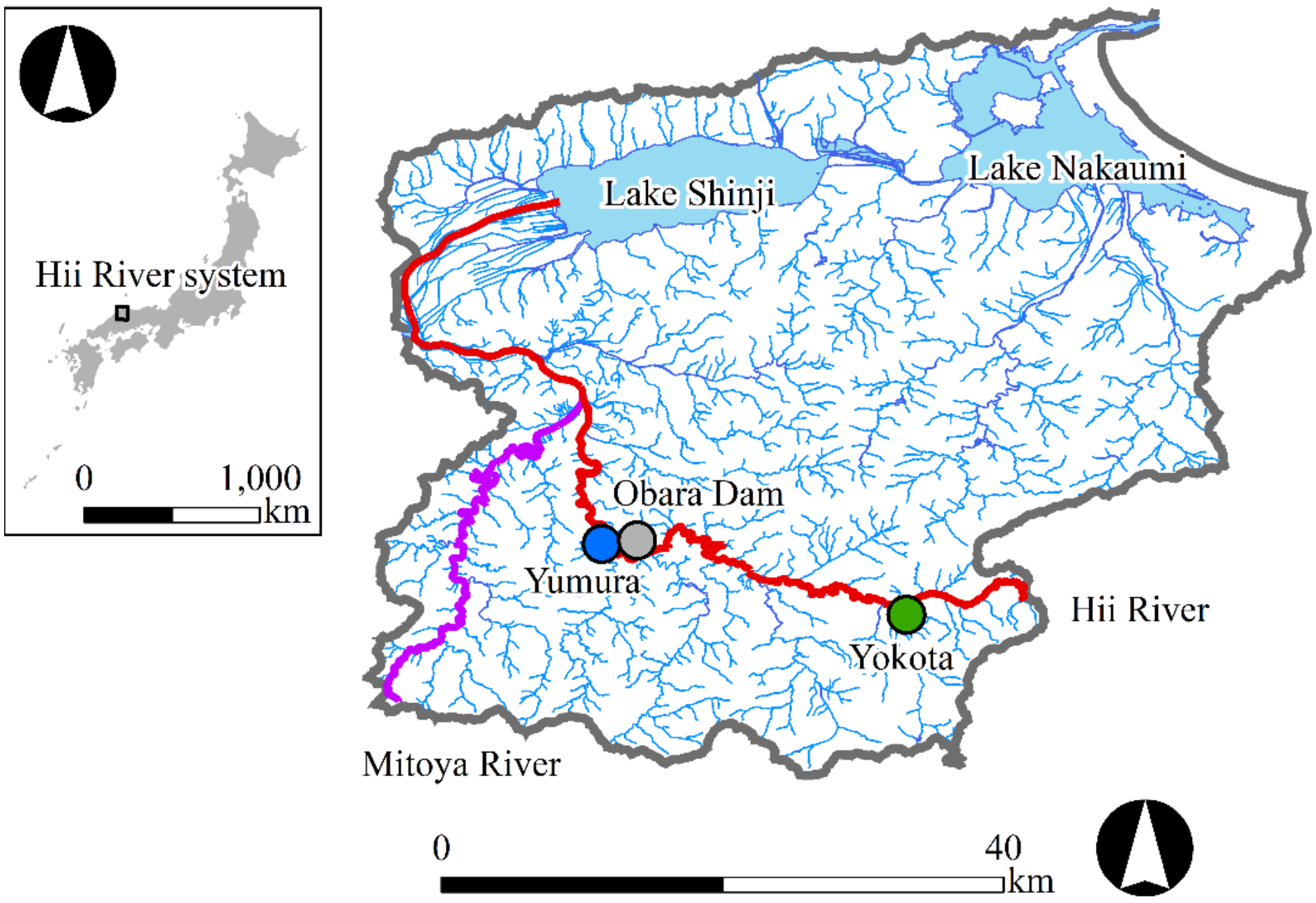

**Figure 1**. The mainstream Hii River and Mitoya River in the Hii River system. Water temperature was measured at Yumura by the authors. Public meteorological records are available at Yokota.

### 3.2 Collected data

The authors have been studying the biology and surrounding water environments of *P. altivelis* since 2015, in cooperation with the Hii River Fisheries Cooperative, which authorizes fishery resources in the mid-to upstream reaches of the Hii River system. The following two data are available: chronological and distributional. Some of the data have been used in past studies investigating fisheries resource management of *P. altivelis* [57]. In particular, the data for 2024 have not been presented elsewhere. The data presented in **Table 3** and **Figures 2-3** are therefore the latest body weight data of *P. altivelis* in the Hii River system at the timing of writing this manuscript. Our models presented later are therefore considered to have higher validity after July 1 on which the harvesting season of the fish opens. The data collected do not include body weights during the spring (March to the end of June) because harvesting the fish is not allowed during this period.

The chronological data contains fishery records from a member of the Hii River Fisheries Cooperative for each year. This member primarily uses toami (casting nets) during the fishing season each year, and records the fish catch, including the total number of fish caught and their total weight, each day. As a result, the chronological data forms time-series averages, although the total number of fish caught varies. We focus on the data in the years 2017, 2018, 2019, 2023, and 2024 because both the chronological and distributional errors are available in these years. **Figure 2** shows that the chronological data fluctuates irregularly over time across all years and quantifying them appears difficult without employing a unified mathematical model, as demonstrated in this study. It should be noted that the different plots in this figure represent different individuals because the fish caught were not released.

The distributional data contains fishery records from approximately 20 members of the Hii River Fisheries Cooperative for one day in each summer from 2017 to 2024, except for the years 2020, 2021, and 2022, when data collection was not possible due to the Covid-19 outbreak in Shimane Prefecture. During each survey, paired members caught *P. altivelis* using a toami for two hours at various locations in the Hii River system (the Hii and Mitoya Rivers), and we recorded the total fish caught and body weight of each fish. Surveys collecting distributional data were conducted on August 6, 2017; August 5, 2018; August 4, 2019; July 30, 2023; and August 25, 2024. **Figure 3** summarizes the collected distributional data. **Table 2** provides the statistics for each survey's distributional data, where $N$ represents the total number of samples. The averages, standard deviation, skewness, and coefficient of variation are abbreviated as Ave, Std, CV, and Skw, respectively. The total number of samples collected could not be controlled, because the surveys were conducted in a natural environment.

The distributional data reveals the empirical PDF of body weight at a single instance. As summarized in **Table 3**, we could sample hundreds of individual fishes for the distributional data in each year. The data indicate that the body weight distribution of *P. altivelis* in the Hii River system exhibits a positive skewness around 1 for all years. The average body weight is around 50 to 55, and the skewness is positive for all the years. The coefficient of variation ranges from 0.3 to 0.4 across all years. The maximum body weight is higher than 100 g, and the minimum is around 20 g. The year 2024 is notably different from the previous five years because it has the smallest average body weight, despite the survey being conducted in most recent season. **Figure 4** illustrates that the fish growth in 2024 appears to be either saturated or terribly slow after the start of the fishing season.

From the chronological and distributional data plotted in **Figures 2-3**, we infer that the lowest growth performance in 2024 is partly due to environmental conditions in the Hii River system. **Table A1** in **Appendix** presents the monthly averages of air temperature in the study area (Yokota, obtained from AMeDAS[2]; see **Figure 1** for the location), showing that the air temperature in 2024 is the highest or ranks high among the years from 2017 to 2024. **Table A2** shows the monthly averages of daily maximum air temperature at Yokota for each month and year, indicating that 2024 is also the highest or ranks high among

[2]https://www.data.jma.go.jp/obd/stats/etrn/ (in Japanese, Last Accessed on November 28, 2024)

the years from 2017 in this measure.

Regarding the water temperature, we acquired 20-min of data (U20, HOBO) in 2022, 2023, and 2024 at Yumura (**Figure 1**), as shown in **Figure A1**. The maximum water temperatures recorded were 29.05 °C in 2022, 29.45 °C in 2023, and 29.95 °C in 2024, indicating a rising trend of 0.4 °C to 0.5 °C per year, with the maximum being highest in 2024. **Figures A2-A3** show the differences in water temperatures between 2024 and 2022, and between 2024 and 2023, respectively. The average differences are 0.716 °C and 1.056 °C for between 2024 and 2022 and between 2024 and 2023, respectively. The positive averages indicate that 2024 was hotter on average. Moreover, the water temperature in 2024 was higher than those in 2022 and 2023 by 71% and 66% of the corresponding measurement period, respectively.

The relatively small water discharge in 2024 is also considered contributed to high water temperature at Yumura. **Figure A4** in **Appendix** shows the water discharge from the Obara Dam[3] at upstream of Yumura. There is no large tributary between the Obara Dam and Yumura. **Figure A4** shows that the flow discharge is lower in most of the time during summer and autumn in 2024 than in 2022 and 2023; the average discharges in 2022, 2023, 2024 from May to October are 9.77 ($m^3$/s), 11.09 ($m^3$/s), and 3.30 ($m^3$/s), respectively. Moreover, we also found that the smallest average discharge in the same period was 2.75 ($m^3$/s) in 2018; however, the air temperature in 2018 was slightly lower than that in 2024 on average (**Tables A1-A2**).

Water temperature in a river is known to positively correlate with or is modelled as an increasing function of air temperature at a nearby site (e.g., [58-60]). Johnson et al. [61] reported that rising river water temperatures can shift ecological regime, leading to lower dissolved oxygen concentrations and negatively affecting fish movement and dispersal. We then infer that the rising water temperatures observed from summer to autumn negatively impacted the growth of *P. altivelis* in the Hii River system.

**Table 3.** Summary statistics of the distributional data: average (Ave), standard deviation (Std), coefficient of variation (CV), skewness (Skw), maximum (Max), and minimum (Min) values.

| | *N* | Ave (g) | Std (g) | CV (-) | Skw (-) | Max (g) | Min (g) |
|---|---|---|---|---|---|---|---|
| 2017 | 234 | 55.6 | 19.1 | 0.344 | 0.78 | 132.0 | 20.5 |
| 2018 | 189 | 57.3 | 18.5 | 0.322 | 1.16 | 152.0 | 16.0 |
| 2019 | 227 | 56.4 | 18.2 | 0.323 | 0.95 | 119.5 | 20.0 |
| 2023 | 297 | 52.2 | 21.0 | 0.402 | 1.42 | 163.0 | 11.0 |
| 2024 | 459 | 49.9 | 15.7 | 0.315 | 1.53 | 127.4 | 13.3 |

[3] Water Information System by MLIT, Japan. http://www1.river.go.jp/cgi-bin/SrchDamData.exe?ID=607041287705020&KIND=1&PAGE=0 (in Japanese, Last Accessed on December 1, 2024)

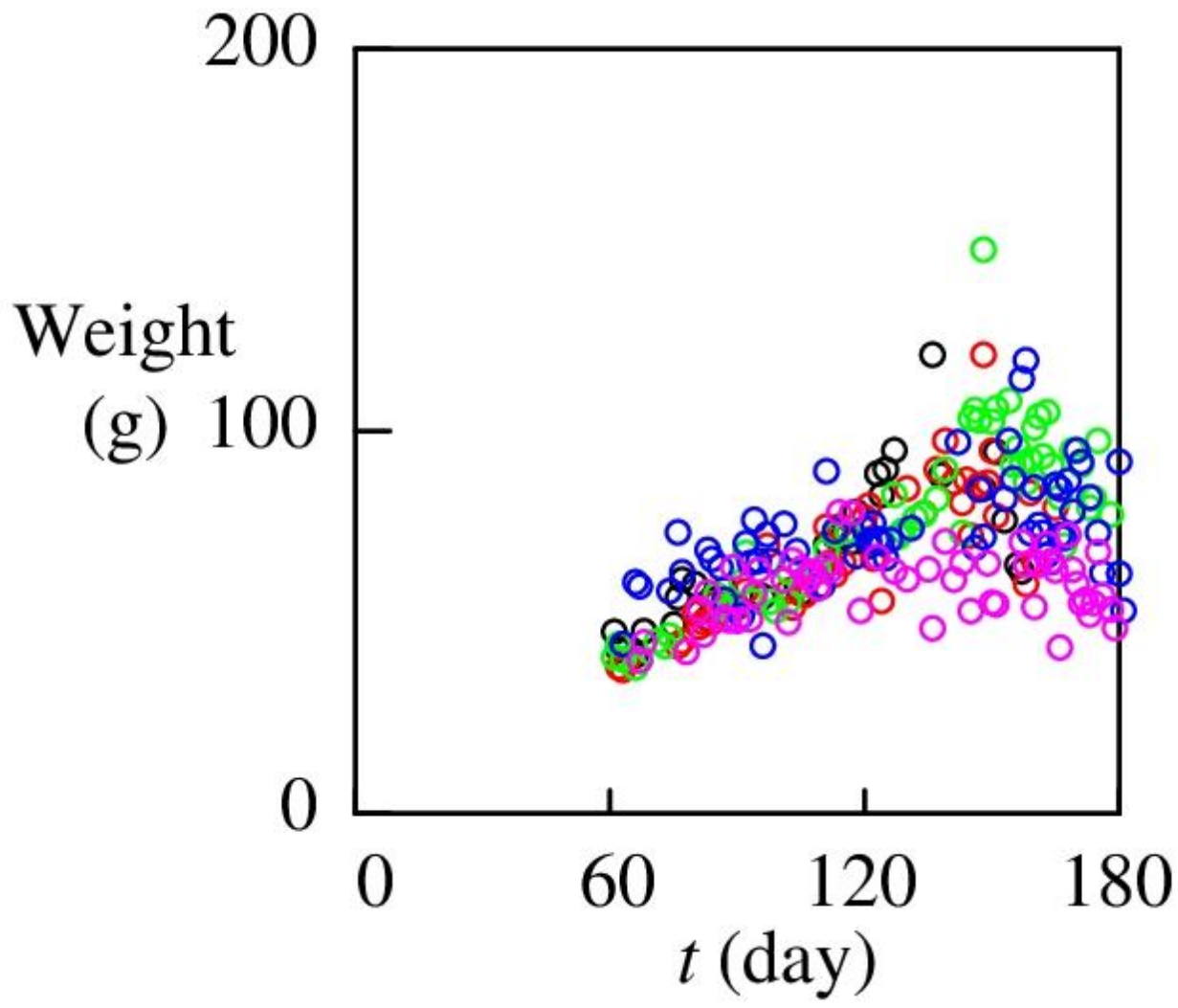

**Figure 2.** Chronological data set of *P. altivelis* in the Hii River in 2017 (black), 2018 (red), 2019 (green), 2023 (blue), and 2024 (magenta).

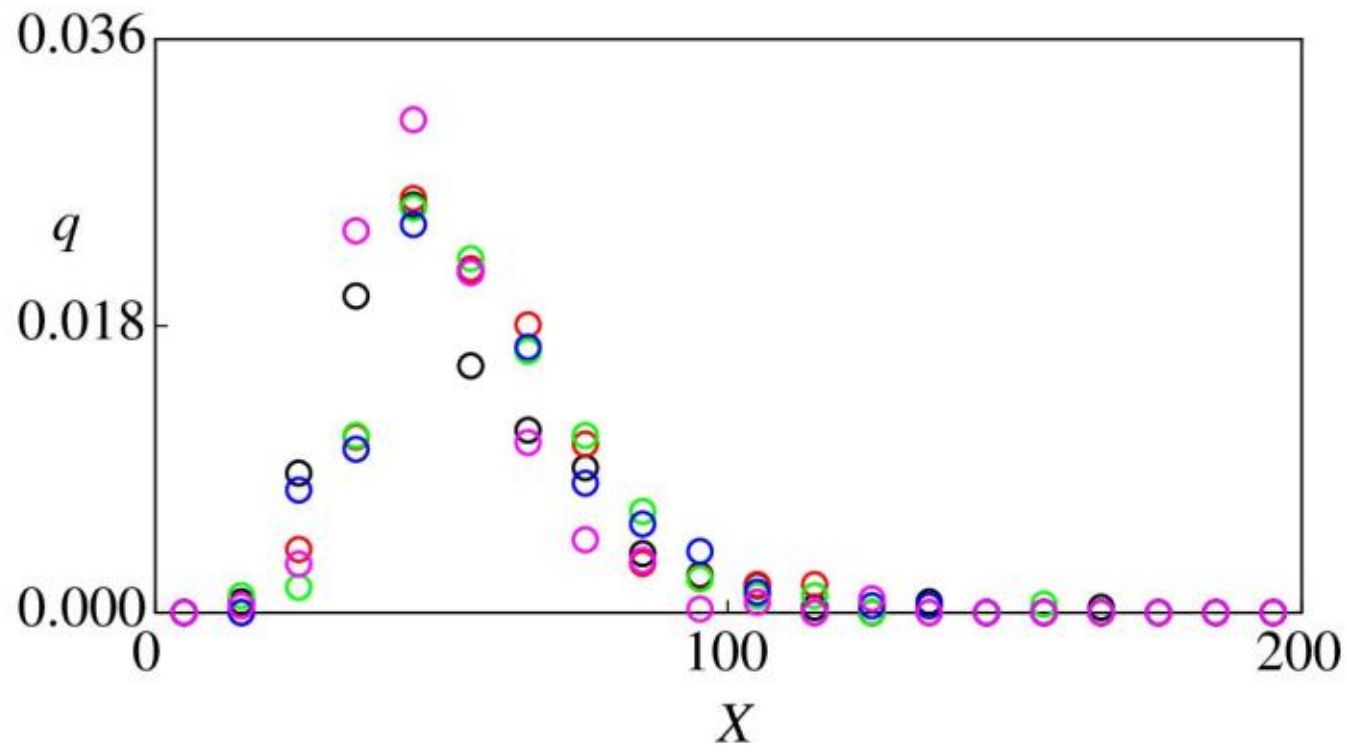

**Figure 3.** Distributional data set of *P. altivelis* in the Hii River in 2017 (black), 2018 (red), 2019 (green), 2023 (blue), and 2024 (magenta). Notice that each data has been obtained on different calendar days: August 6 in 2017, August 5 in 2018, August 4 in 2019, July 30 in 2023, and August 25 in 2024.

## 4. Application

### 4.1 Model identification

We find the parameters $u, r, \alpha, \beta$ for both the Gamma and inverse Gamma cases. The identification procedure of the proposed model is as follows: the time of the survey for collecting the distributional data is denoted as $t_c$, with the initial time being set as the beginning of May 1 without any loss of generality. The subscript "emp" refers to the empirical value. Skewness is not used for identifying parameter values but will serve as a statistic to validate identified models. We only have weight data of *P. altivelis* in the Hii River system but not the corresponding length data. In this application, we therefore regard the variable $L$ as an auxiliary one to explain the temporal variation of the PDF of body weight.

**Identification procedure**

**0.** Prepare the chronological and distributional data for a year.

**1.** Identify the quantities $\alpha$ and $\beta L_{t_c}^3$ by using moment matching:

**(Gamma case)** $$\alpha = \left.\frac{\mathrm{Ave}_{\mathrm{emp}}^2}{\mathrm{Std}_{\mathrm{emp}}^2}\right|_{t=t_c} \text{ and } \beta = \left.\frac{\mathrm{Std}_{\mathrm{emp}}^2}{\mathrm{Ave}_{\mathrm{emp}}}\right|_{t=t_c} L_{t_c}^{-3} \tag{35}$$

and

**(Inverse Gamma case)** $$\alpha = 2 + \left.\frac{\mathrm{Ave}_{\mathrm{emp}}^2}{\mathrm{Std}_{\mathrm{emp}}^2}\right|_{t=t_c} \text{ and } \beta = \left.\left[\mathrm{Ave}_{\mathrm{emp}}\left(1+\frac{\mathrm{Ave}_{\mathrm{emp}}^2}{\mathrm{Std}_{\mathrm{emp}}^2}\right)\right]\right|_{t=t_c} L_{t_c}^{-3}. \tag{36}$$

The parameter $\alpha$ is identified at this step. Here, $L_{t_c}$ contains the two parameters $u,r$.

**2.** Rewrite the average body weight as

$$\mathbb{E}[X_t] = \mathbb{E}[K]L_t^3 = \left(\mathbb{E}[K]L_{t_c}^3\right)\frac{L_t^3}{L_{t_c}^3} = \mathrm{Ave}_{\mathrm{emp}}\big|_{t=t_c}\frac{L_t^3}{L_{t_c}^3}. \tag{37}$$

In the right-hand side of (37), unknown parameters are $u,r$.

**3.** Solve the following minimization problem using a common nonlinear least-squares method to identify the values of $u,r$:

$$\min_{u,r\geq 0}\sum_{j=1}^{N}\left(E\left[X_{t_j}\right]-\bar{X}_{\mathrm{emp},t_j}\right)^2. \tag{38}$$

At the end of this step, the values of the parameters $\alpha,u,r$ are available.

**4.** Identify the value of $\beta$ by reusing the second equation of (35) or (36).

**Table 4** shows the identified parameter values for the Gamma case for each year. Similarly, **Table 5** presents the identified parameter values for the inverse Gamma case for each year. **Table 6** compares empirical and modelled skewness. **Figure 4** compares the empirical and modelled PDFs of the distributional data, and **Figure 5** compares the modelled average and standard deviations with the chronological data. According to **Figure 4**, both the Gamma and inverse Gamma cases capture the empirical PDFs, and the latter better captures their modes. **Table 6** suggests that the Gamma case underestimates the empirical skewness, while the inverse Gamma case overestimates it, partly because of the heavier tail of the latter. The relative errors between the empirical and modelled skewness are mostly at the order of $O\left(10^{-1}\right)$, and the Inverse Gamma case has a larger range of the relative errors than the Gamma one, varying from 0.085 in 2024 to 1.000 in 2017. The fitted model shows that, for each year, 2024 has the lowest performance on average during summer to autumn (i.e., approximately 90 to 180 days) among the four years examined in this study. **Tables 4-5** suggest that the year 2024 is characterized by a small growth rate $r$; and the resulting average growth curve in **Figure 5(e)** implies that the growth speed of body weight is indeed slower than that of the other years. The agreement of the model with the distributional data, as shown in **Figure 4(e)**, supports the model's validity, whereas the fitted parameter values imply that the growth regime in 2024 is qualitatively different from those of the other years. Nevertheless, this study has a limitation that there is a lack of linkages between the identified parameter values and environmental conditions such as water temperature and discharge. Exploring this point will be addressed in our future research along with further field investigations because the main objective of this study is to propose and apply the model of **Section 2**.

**Table 4.** Identified parameter values of the proposed model: Gamma case.

| | $\alpha$ (-) | $\beta$ (g) | $u$ (-) | $r$ (1/day) |
|---|---|---|---|---|
| 2017 | 8.47.E+00 | 5.28.E+01 | 3.22.E-01 | 3.12.E-03 |
| 2018 | 9.59.E+00 | 1.16.E+01 | 2.20.E-01 | 1.42.E-02 |
| 2019 | 9.60.E+00 | 1.37.E+01 | 1.84.E-01 | 1.26.E-02 |
| 2023 | 6.18.E+00 | 1.63.E+01 | 3.11.E-01 | 1.39.E-02 |
| 2024 | 1.01.E+01 | 6.16.E+06 | 8.21.E-03 | 9.43E-06 |

**Table 5.** Identified parameter values of the proposed model: inverse Gamma case.

| | $\alpha$ (-) | $\beta$ (g) | $u$ (-) | $r$ (1/day) |
|---|---|---|---|---|
| 2017 | 1.05.E+01 | 4.24.E+03 | | |
| 2018 | 1.16.E+01 | 1.18.E+03 | | |
| 2019 | 1.16.E+01 | 1.40.E+03 | Same with **Table 4** | |
| 2023 | 8.18.E+00 | 7.24.E+02 | | |
| 2024 | 1.21.E+01 | 6.91.E+08 | | |

**Table 6.** Empirical and modelled skewness compared to distributional data. The number inside "( )" represents the relative error between the empirical and modelled skewness.

| | Empirical | Gamma case | Inverse Gamma case |
|---|---|---|---|
| 2017 | 0.78 | 0.69 (0.115) | 1.56 (1.000) |
| 2018 | 1.16 | 0.65 (0.440) | 1.44 (0.241) |
| 2019 | 0.95 | 0.65 (0.316) | 1.44 (0.516) |
| 2023 | 1.42 | 0.81 (0.430) | 1.92 (0.352) |
| 2024 | 1.53 | 0.63 (0.588) | 1.40 (0.085) |

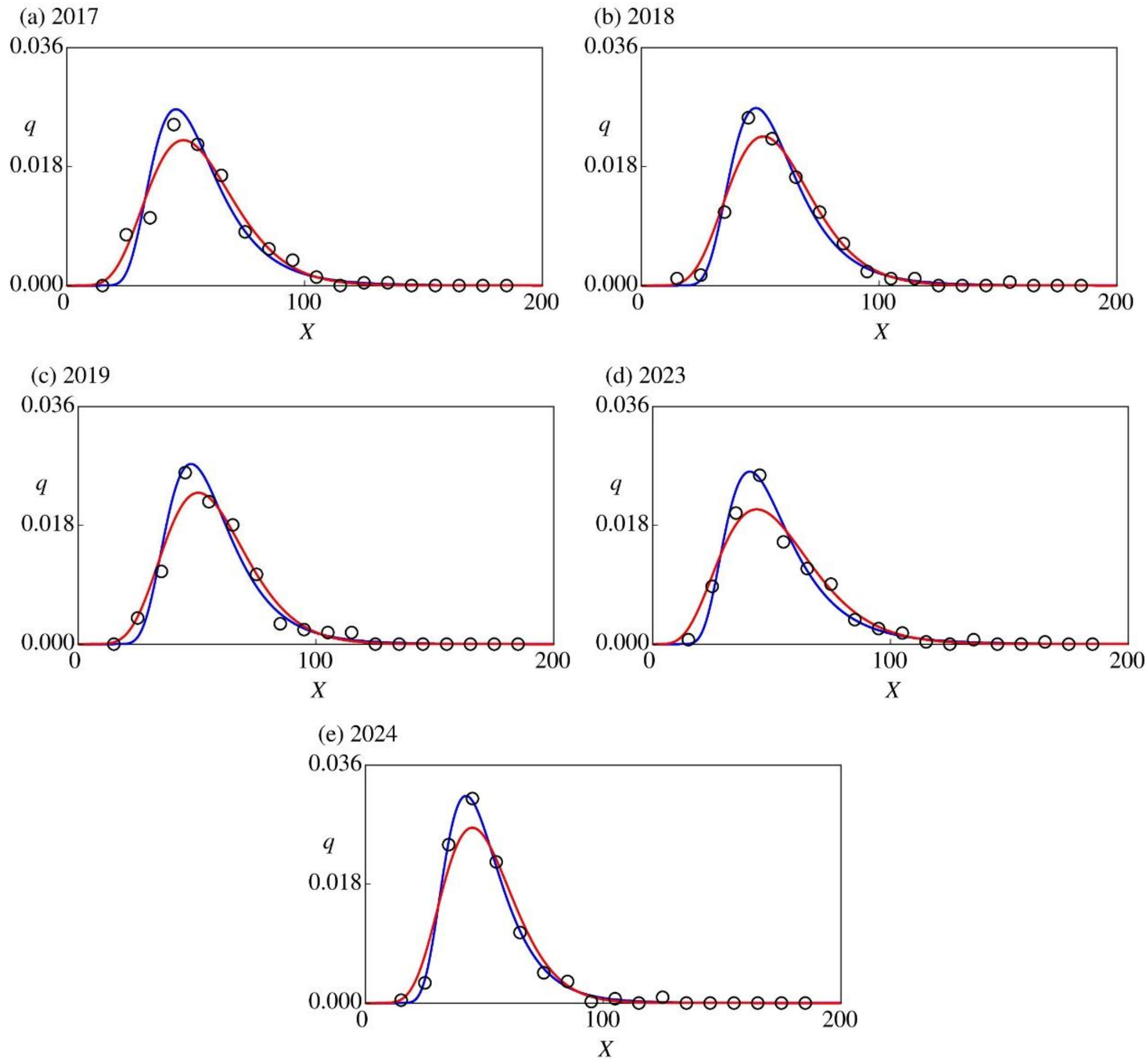

**Figure4.** Comparison of the PDFs $q$ of the body weight $X$ : fitted model with the Gamma-type $p$ (red curve), fitted model with the inverse Gamma-type $p$ (blue curve), and distributional data (black circle) in each year.

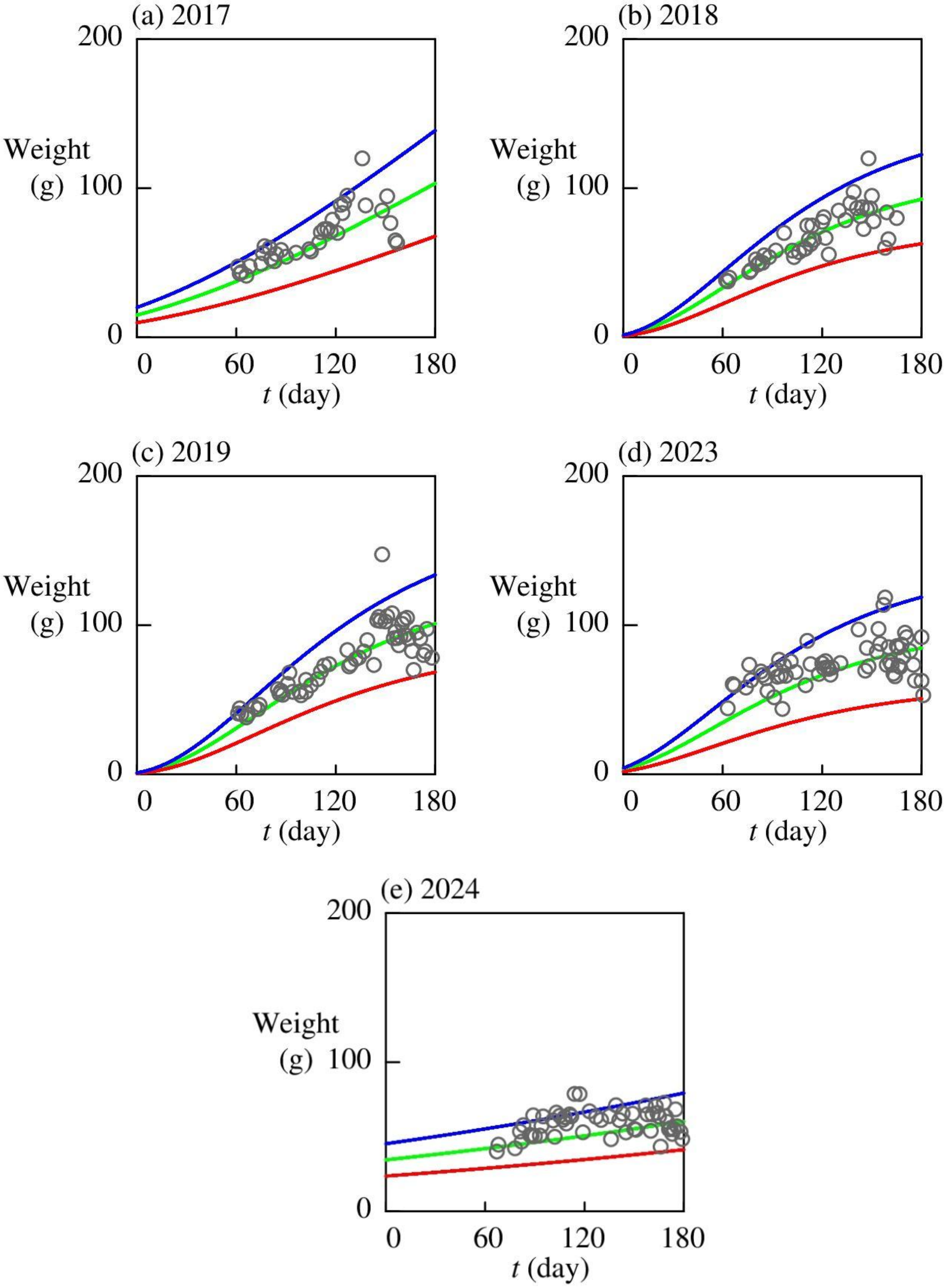

**Figure5.** Comparison of the empirical and modelled body weights in each year. Both the Gamma and inverse Gamma cases give the same average and standard deviation of the body weight due to the moment matching-based identification procedure: empirical data (circles), theoretical average minus standard deviation (red), theoretical average (green), theoretical average plus standard deviation (blue).

### 4.2 Influences of model misspecification

According to **Table 3**, for the Gamma case, for each given $\eta$ a model with a higher value of $\beta$ is more sensitive to misspecifications. In this view, as shown in **Table 4**, the most sensitive model, except for the year 2024, is that of 2017, while the least sensitive is that of 2018. The year 2024 stands out among all the examined years due to an exceptionally large $\beta$. **Figure 6** illustrates the coefficient of multiplication, denoted as $\mathrm{Coe}$, as a function of $\eta$ for (a) the upper-bound case ($\mathrm{Coe} = \frac{1}{1-\beta/\eta}$) and (b) lower-bound case ($\mathrm{Coe} = \frac{1}{1+\beta/\eta}$) in each year. Similarly, **Figure 7** displays the coefficient of multiplication $\mathrm{Coe}$ as a function of $\varepsilon$. The $\eta$-based formulation of $\mathrm{Coe}$ demonstrates how the modeler's uncertainty aversion affects the upper- and lower-bounds. Additionally, the $\varepsilon$-based formulation highlights how these bounds should be determined based on the assumed error bound.

In practice, we must estimate a reasonable range of $\varepsilon$ or $\eta$ to quantify the influence of possible misspecifications. Because the average body weight of the Gamma case is proportional to $\beta$ and the distorted PDFs are given by (14) and (23), one possible way is to firstly assume the relative error $\delta$ of the estimated $\beta$. More specifically, under the worst-case scenario in the sense of relative entropy, by the replacement $\beta \to \beta(1 \pm \delta)$ in **Table 1** under the assumption that the average (and hence the variance) of a misspecified model is given according to **Table 2**, we obtain

$$\alpha\beta(1 \pm \delta) = \frac{\alpha\beta}{1 \mp \beta/\eta} \quad \text{or equivalently} \quad \eta = \pm\beta\left(1 - \frac{1}{1 \pm \delta}\right)^{-1}. \tag{39}$$

Regarding "$\pm$" in the right-hand equation of (39), the plus and minus signs should be taken for the upper- and lower-bounding cases, respectively. The equation (39) connects the relative error $\delta$ to the uncertainty aversion $\eta$. We can also estimate the corresponding $\varepsilon$ using (18) or (27) once an $\eta$ is given. **Table 7** presents the error bound $\varepsilon$ as a function of the relative error $\delta$ for each year, suggesting that $\varepsilon$ can be assumed to be at most $O(10^0)$ for $\delta$ in the range of a few tens of percent.

Finally, modeling body weight is not only important by itself but is also useful for estimating the number of eggs per female individual of *P. altivelis*. Uchida et al. [62] discussed that each female lays approximately 800 eggs per gram of body weight, suggesting that the total number of eggs laid by a single female is proportional to its body weight $X$. Using this estimation, we calculate the total number of eggs per female in each year. In this view, the average total number of eggs laid by one female under model uncertainty with a prescribed $\eta$ is $800(1+\beta/\eta)^{-1}\mathbb{E}[X_{180}]$ for the lower-bounding case and $800(1-\beta/\eta)^{-1}\mathbb{E}[X_{180}]$ ($\eta > \beta$) for the upper-bounding case. Therefore, the proposed misspecification framework can be directly incorporated into the estimation of the number of eggs laid.

We assume that the spawning event occurs each year from the middle of October to the early

November[4], corresponding to approximately day 180 in our context. **Figure 8** compares the average body weight across years, showing that the number of eggs per female was lowest in 2024, followed by 2023, 2018, 2017, and 2019. Although further discussion should consider population dynamics, this type of biological analysis under uncertain conditions is useful for quantifying the life histories of the fish species. In particular, the closed-form estimates of eggs per individual are significant because the data revealed approximately 1.5-fold difference between the smallest (2024) and largest (2017) years; according to the proposed model, the average body weight of individual *P. altivelis* at the end of September 2024 is 0.64, 0.66, 0.62, and 0.72 times of those in 2017, 2018, 2019, and 2023, respectively. Further rising water temperature of the river water would more seriously affect the growth and population dynamics of *P. altivelis* in the Hii River system.

**Table 7.** The error bound $\varepsilon$ as a function of $\delta$ for each year.

| | $\delta$ | 2017 | 2018 | 2019 | 2023 | 2024 |
|---|---|---|---|---|---|---|
| Upper-bound | −0.3 | 1.61.E+00 | 1.83.E+00 | 1.83.E+00 | 1.18.E+00 | 1.92.E+00 |
| | −0.2 | 3.87.E-01 | 4.38.E-01 | 4.38.E-01 | 2.82.E-01 | 4.61.E-01 |
| | −0.1 | 6.12.E-02 | 6.92.E-02 | 6.93.E-02 | 4.46.E-02 | 7.29.E-02 |
| Lower-bound | 0.1 | 3.12.E-02 | 3.53.E-02 | 3.53.E-02 | 2.27.E-02 | 3.72.E-02 |
| | 0.2 | 9.57.E-02 | 1.08.E-01 | 1.08.E-01 | 6.98.E-02 | 1.14.E-01 |
| | 0.3 | 1.71.E-01 | 1.93.E-01 | 1.93.E-01 | 1.24.E-01 | 2.03.E-01 |

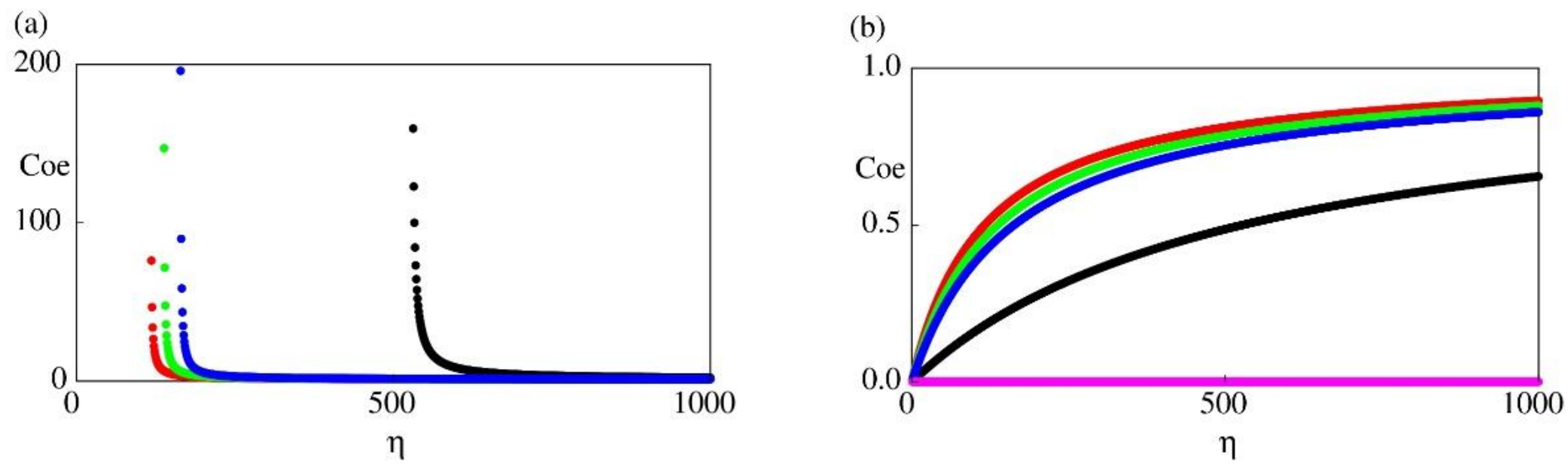

**Figure 6.** The coefficient of multiplication Coe as a function of $\eta$ for (a) the upper-bounding ($\mathrm{Coe} = \dfrac{1}{1-\beta/\eta}$) and (b) lower-bounding cases ($\mathrm{Coe} = \dfrac{1}{1+\beta/\eta}$) in 2017 (black), 2018 (red), 2019 (green), 2023 (blue), and 2024 (magenta). The results for 2024 are not provided in **Figure 6(a)** because of the large $\beta$ with which Coe becomes negative values close to 0.

[4] Report on the FY2021 project to develop a methodology for estimating the carrying capacity of the environment, Shimane Prefecture. Available at: https://www.maff.go.jp/j/budget/yosan_kansi/sikkou/tokutei_keihi/seika_R03/attach/pdf/itaku_R03_ippan-465.pdf (in Japanese; last accessed November 27, 2024)

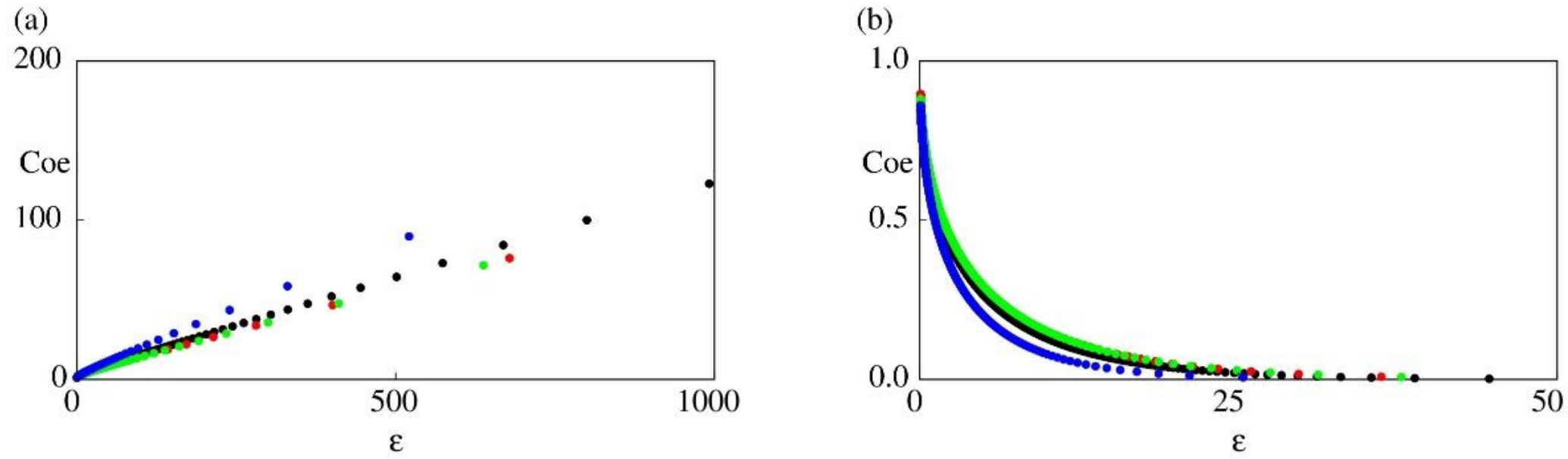

**Figure 7.** The coefficient of multiplication Coe as a function of $\varepsilon$ for (a) the upper-bounding ( $\mathrm{Coe}=\dfrac{1}{1-\beta/\eta}$ ) and (b) lower-bounding cases ( $\mathrm{Coe}=\dfrac{1}{1+\beta/\eta}$ ) in 2017 (black), 2018 (red), 2019 (green), 2023 (blue). The results for 2024 were not plotted in **Figures 7(a)-(b);** the plots are concentrated at the lower left in the panel of **Figure 7(a)** and Coe is between 100 and 180 (very large) in **Figure 7(b)**.

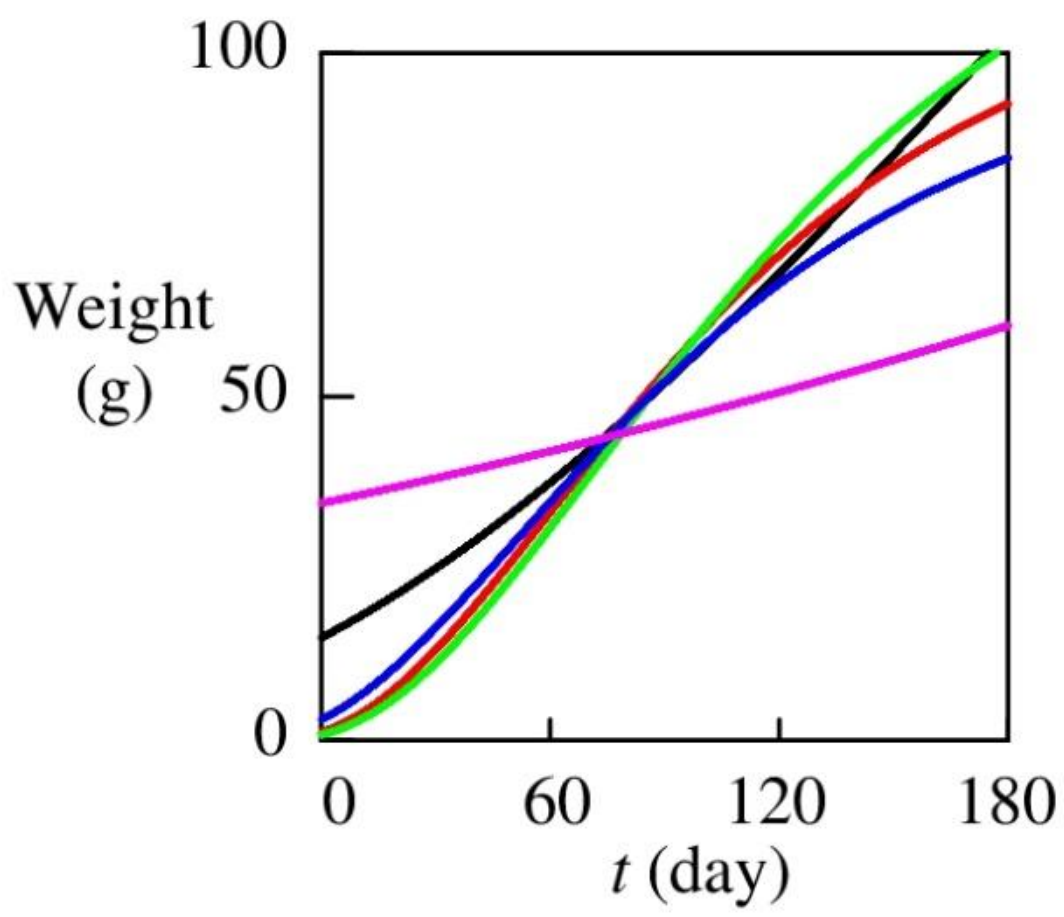

**Figure 8.** Comparison of the average body weight of *P. altivelis* in the Hii River in 2017 (black), 2018 (red), 2019 (green), 2023 (blue), and 2024 (magenta). The number of eggs is simply proportional to the average body weight as explained in **Section 4.2**. The vertical axis of the figure can therefore be understood as a scaled version of the number of eggs.

### 4.3 Influences of stochastic fluctuation

We briefly examined the SDE (33) for the collected data of *P. altivelis*. An additional parameter, $\sigma$, needs to be identified. We identify the model parameters by simply replacing each $L_t^3$ in the **identification procedure** with its expectation $M_t^{(3)}$ (see **Appendix**) and using a similar nonlinear least-squares method, considering the expectation of $\sigma$. The identified results are presented in **Table A3** of **Appendix** for the Gamma case, indicating that $\sigma^2$ is several orders of magnitude smaller than $r$. In this view, the stochastic fluctuation does not explain the small body weight in the year 2024 as shown in **Table A3**. This identification result implies that accounting for individually different environmental fluctuations may not have been effective for *P. altivelis* in the Hii River system in the years studied.

A more realistic scenario would involve employing stochastic fluctuations that are not entirely independent among individuals but exhibit certain correlation. However, identifying such correlations would be challenging because it is technically difficult to track the growth of individual fish in their natural environment. While using a more complex model could theoretically resolve this issue, it would sacrifice the analytical tractability of the proposed model. This is because the SDE (33) has been designed as a minimal stochastic extension of the proposed model. This aspect will be explored in future studies. Nonetheless, the current study suggests that simply adding the effects of the physiological heterogeneity and stochastic fluctuation may not be sufficient for accurately studying the biological growth of the fish.

Finally, we investigate the sensitivity of the stochastic version of the proposed model to fluctuations to better understand the role of stochasticity, which may be relevant for data not covered in this study. We identified the model with stochastic fluctuations for the prescribed values of $\sigma$, as shown in **Tables A3-A6** in **Appendix**. For the case study, we select the parameters for the year 2023 using the Gamma PDF (**Table 3**). The average and standard deviation are then computed by using the moment formula (40). **Figure 9** shows the time evolution of the average, standard deviation, and skewness for the identified model with and without stochastic fluctuations. The average body weight converges to the same value at time 90 days, which is when the distributional data are collected in 2023. The results of $\sigma = 0$ (1/day$^{1/2}$) and $\sigma = 0.01$ (1/day$^{1/2}$) are comparable. However, when $\sigma = \sqrt{0.001}$ (1/day$^{1/2}$), the identified model with stochasticity predicts smaller average body weights before day 90 and larger average body weights after day 90 (**Figure 9(a)**). Interestingly, assuming the presence of stochasticity enhances growth during the later seasons. The standard deviation is larger for models with higher prescribed $\sigma$, except near the time 0 (**Figure 9(b)**). Thus, the body growth dynamics are predicted to result in larger average body weights and greater individual differences in the later season, approximately from August to October in this case. Finally, the presence of stochasticity appears to monotonically increase the skewness (**Figure 9(c)**). This finding is significant because the model without stochasticity underestimates the observed skewness, as shown in **Table 6**. Incorporating a proper mechanism for stochastic fluctuations could enhance the performance of the proposed model. This aspect will be explored in our future research.

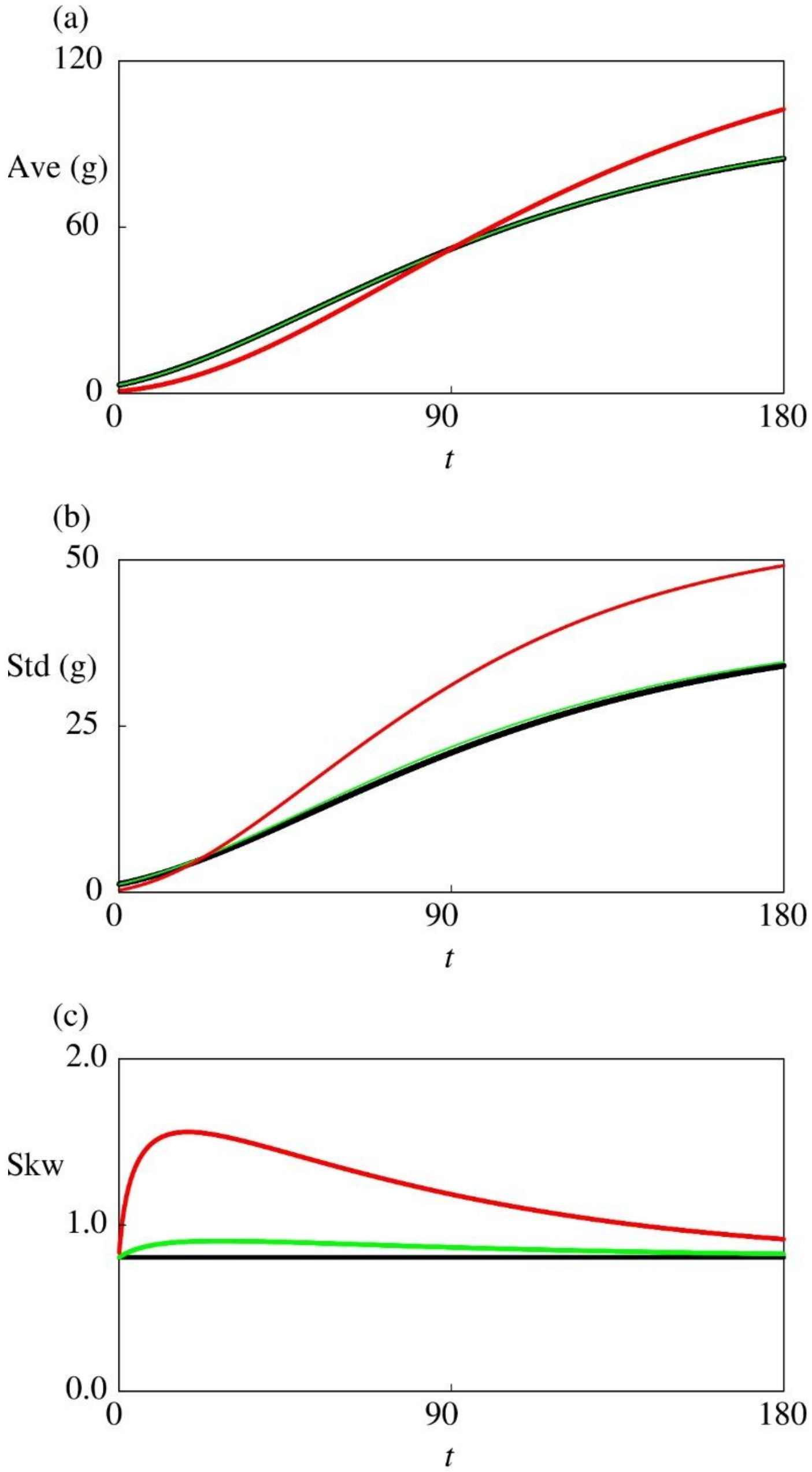

**Figure 9.** Statistics of the proposed model with stochastic fluctuation at each time $t$ (day): (a) Average, (b) Standard deviation, and (c) Skewness. Each color represents the model without stochastic fluctuation (black), model with $\sigma = 0.01$ (1/day$^{1/2}$) (green), and model with $\sigma = \sqrt{0.001}$ (1/day$^{1/2}$) (red).

## 5. Conclusions

We proposed a fully analytical framework for modeling the body weight dynamics of individual fish within a physiologically heterogeneous population. The model was based on a VB model with a distributed maximum body weight, incorporating potential misspecifications through the relative entropy between the true and identified models. The proposed model was applied to the body weight data collected for *P. altivelis* in the Hii River, Japan. Characteristics of the data were discussed based on the temperature and discharge data in and around the river. We evaluated the model using Gamma and inverse Gamma PDFs for the distribution of maximum body weights derived from the data. Our results demonstrated that both cases correctly captured the average, standard deviation, and positive skewness of the body weight distributions. Notably, the model based on the inverse Gamma PDF of the maximum body weight predicted a larger skewness in the body weight distribution and a sharper peak compared to the Gamma model. Influences of the misspecifications were also investigated for the Gamma case with an emphasis of the reproduction of the fish.

This study did not address several key issues. For example, the biological growth of an individual fish depends on surrounding environmental conditions, such as water temperature [63]. Model parameters in the proposed framework should be parameterized through environmental variables in future studies. Another limitation of the model is that it applies only to the body weight data collected from summer to autumn, leaving its applicability to the larval and juvenile stages of the fish uncertain. Furthermore, the model applies only to fish species without an age structure and require customization for age-structured populations. From a biological perspective, employing an agent-based model or its macroscopic mean-field version could provide a more mechanistic description of biological growth phenomena.

We conjecture that a more reasonable approach would be to account for physiological heterogeneity and environmental fluctuations in a more interactive manner, because the latter may influence the former. One strategy is to incorporate both independent and common-noise processes among individuals. It has also noted that the flow discharge in the Hii River system shows long memory, which is more complex than the Brownian motion employed in this study [64]. Integrating environmental data in the proposed growth model would open the door to more advanced stochastic models for fish population dynamics. Modeling body weight is also important for the evaluation of carbon sequestration in aquatic environment [65] and nutrient cycling between upstream and downstream river reaches due to migratory behavior and body wastes, as reported inland fishes including *P. altivelis* [66]. These issues will be addressed in future.

## Appendices

### A.1 Moments of Jacobi process

The moment $M_t^{(m)} = \mathbb{E}\left[\left(L_t^{(i)}\right)^m\right]$ ($m = 1, 2, 3, \ldots$), which is common to all individuals, can be found in the following system of ODEs

$$\frac{\mathrm{d}}{\mathrm{d}t} M_t^{(m)} = \left( mr + \frac{1}{2} m(m-1)\sigma^2 \right)\left( M_t^{(m-1)} - M_t^{(m)} \right), \quad t > 0 \quad \text{and} \quad m = 1, 2, 3, \ldots \tag{40}$$

subject to the initial condition $M_0^{(m)} = u^m$ ($m = 0, 1, 2, \ldots$). Here, $M_t^{(0)} = 1$ ($t \geq 0$).

We can obtain an exact solution to the system by (long) analytical calculations. The first three moments are presented as follows: For $t \geq 0$,

$$M_t^{(1)} = 1 - (1-u)e^{-rt}, \tag{41}$$

$$M_t^{(2)} = u^2 e^{-(2r+\sigma^2)t} + 1 - e^{-(2r+\sigma^2)t} - \frac{2r+\sigma^2}{r+\sigma^2}(1-u)\left(e^{-rt} - e^{-(2r+\sigma^2)t}\right), \tag{42}$$

and

$$\begin{aligned} M_t^{(3)} = {} & u^3 e^{-(3r+3\sigma^2)t} + 1 - e^{-(3r+3\sigma^2)t} + \frac{u^2(3r+3\sigma^2)}{r+2\sigma^2}\left(e^{-(2r+\sigma^2)t} - e^{-(3r+3\sigma^2)t}\right) \\ & - \frac{3r+3\sigma^2}{r+2\sigma^2}\left(e^{-(2r+\sigma^2)t} - e^{-(3r+3\sigma^2)t}\right) \\ & - \frac{(2r+\sigma^2)(3r+3\sigma^2)}{r+\sigma^2}(1-u)\left\{ \frac{1}{2r+3\sigma^2}\left(e^{-rt} - e^{-(3r+3\sigma^2)t}\right) - \frac{1}{r+2\sigma^2}\left(e^{-(2r+\sigma^2)t} - e^{-(3r+3\sigma^2)t}\right) \right\}. \end{aligned} \tag{43}$$

**A.2 Monthly air temperature in the study area**

**Table A1** presents the monthly air temperatures recorded at the Yokota AMeDAS station since 2017. **Table A2** lists the monthly averages of the daily maximum air temperature at Yokota. **Figure A1** shows the water temperature at Yumura for the years 2022, 2023, and 2024. **Figure A2** shows the difference in water temperatures at Yumura between 2024-2022. Similarly, **Figure A3** shows the difference in water temperatures at Yumura between 2024-2022. **Figure 4** plots water discharge ($m^3$/s) from the Obara Dam in 2022 (green), 2023 (blue), and 2024 (red), where the data has been taken from Water Information System by MLIT, Japan.

**Table A1.** Monthly averages of air temperature in degrees Celsius (°C) at Yokota. **Red values** represent the highest temperature for each month over the years. **Purple values** represent the second highest value.

| | 2017 | 2018 | 2019 | 2020 | 2021 | 2022 | 2023 | 2024 |
|---|---|---|---|---|---|---|---|---|
| January | 0.8 | -0.6 | 1.2 | 3.7 | -0.2 | 0.3 | 0.8 | 2.2 |
| February | 1.0 | -0.7 | 3.2 | 3.3 | 3.4 | -0.6 | 2.2 | 3.9 |
| March | 3.7 | 6.7 | 5.6 | 6.5 | 7.6 | 6.7 | 8.3 | 5.4 |
| April | 11.8 | 12.3 | 9.7 | 8.2 | 10.8 | 12.0 | 11.8 | 14.0 |
| May | 16.8 | 16.5 | 16.6 | 16.6 | 16.2 | 15.8 | 16.3 | 16.2 |
| June | 18.4 | 19.9 | 19.5 | 21.3 | 20.2 | 21.2 | 20.7 | 21.0 |
| July | 25.1 | 25.8 | 23.3 | 22.4 | 24.6 | 24.7 | 25.4 | 26.1 |
| August | 24.7 | 25.5 | 25.3 | 26.2 | 24.1 | 25.5 | 26.7 | 26.4 |
| September | 18.8 | 19.8 | 21.7 | 20.5 | 20.7 | 21.6 | 22.7 | 24.0 |
| October | 14.4 | 13.8 | 15.4 | 13.1 | 14.9 | 13.8 | 13.5 | 16.7 |

**Table A2.** Monthly averages of daily maximum air temperature in degrees Celsius (°C) at Yokota. **Red values** represent the highest temperature for each month over the years. **Purple values** represent the second highest value.

| | 2017 | 2018 | 2019 | 2020 | 2021 | 2022 | 2023 | 2024 |
|---|---|---|---|---|---|---|---|---|
| January | 5.1 | 3.6 | 6.0 | 7.9 | 4.0 | 4.7 | 5.6 | 6.6 |
| February | 5.7 | 4.5 | 8.4 | 8.8 | 9.8 | 4.0 | 7.8 | 8.6 |
| March | 9.8 | 14.3 | 11.7 | 12.8 | 14.8 | 13.5 | 15.9 | 10.6 |
| April | 18.2 | 19.9 | 16.5 | 14.9 | 18.5 | 19.9 | 18.6 | 21.4 |
| May | 24.1 | 22.5 | 24.4 | 22.9 | 22.3 | 23.4 | 23.2 | 23.2 |
| June | 24.8 | 25.4 | 25.0 | 27.3 | 26.1 | 27.2 | 26.5 | 26.9 |
| July | 30.5 | 31.5 | 28.2 | 26.3 | 30.4 | 30.1 | 31.0 | 31.1 |
| August | 30.0 | 31.7 | 30.8 | 32.7 | 29.3 | 31.1 | 32.2 | 33.0 |
| September | 24.4 | 23.9 | 27.3 | 25.8 | 25.5 | 26.9 | 28.0 | 30.5 |
| October | 18.9 | 19.2 | 20.6 | 19.3 | 21.3 | 20.2 | 20.4 | 22.2 |

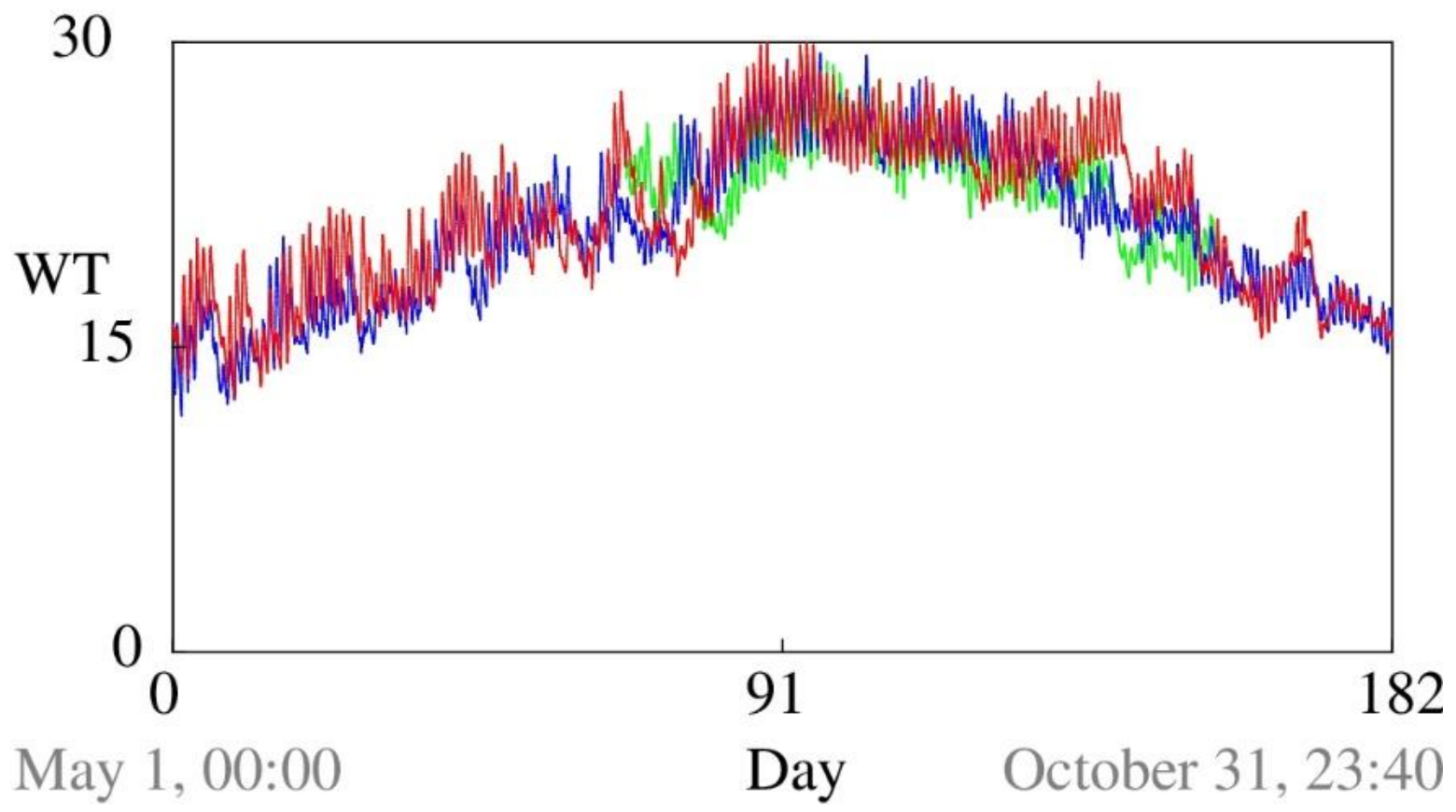

**Figure A1.** Water temperature (WT) at Yumura for the years 2022 (green), 2023 (blue), and 2024 (red). The x-axis represents the time in days, while the y-axis represents the water temperature in degrees Celsius (°C).

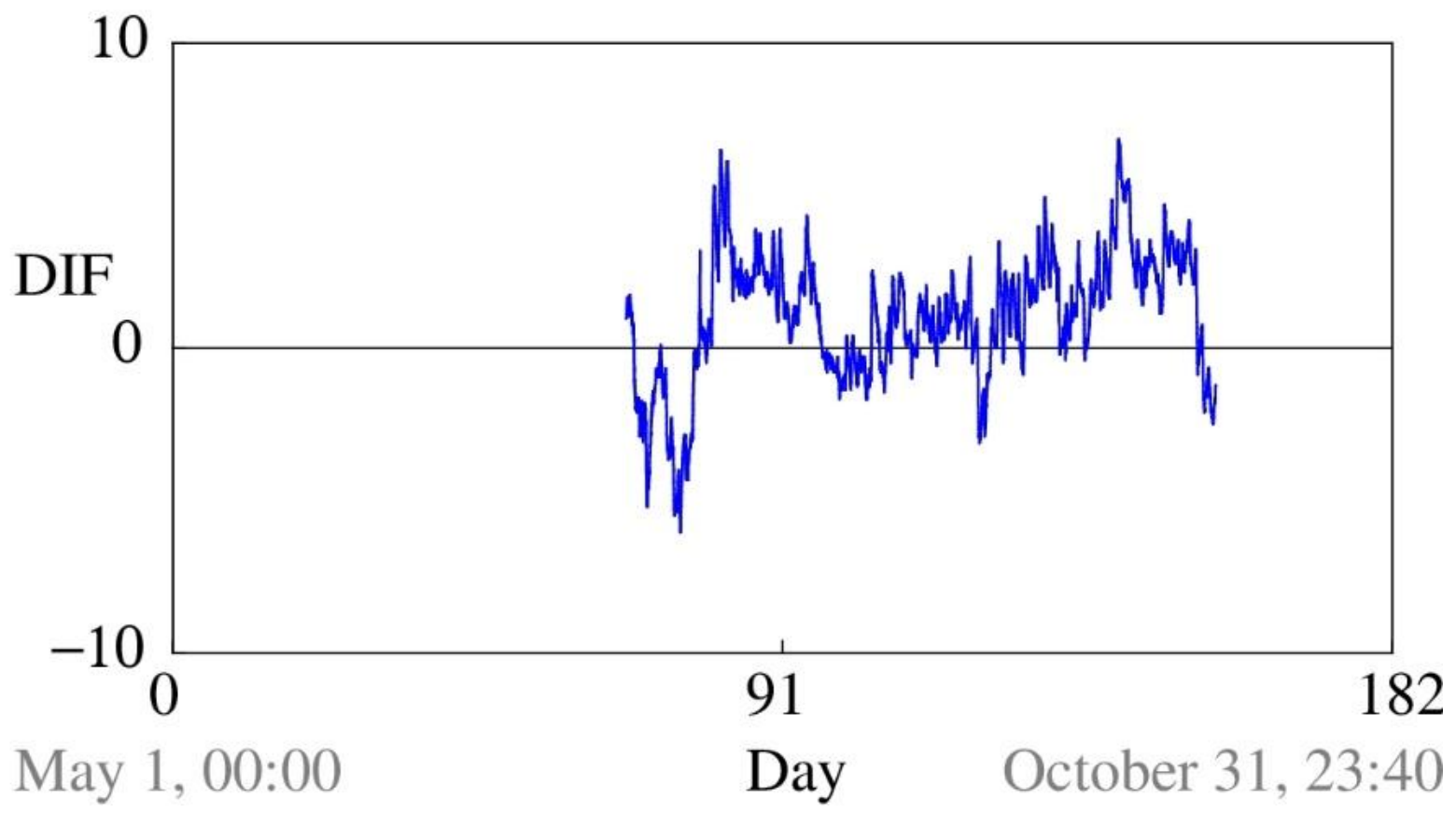

**Figures A2.** Difference (DIF) in water temperatures at Yumura between 2024-2022. The x-axis represents the time in days, while the y-axis represents the temperature difference in degrees Celsius (°C).

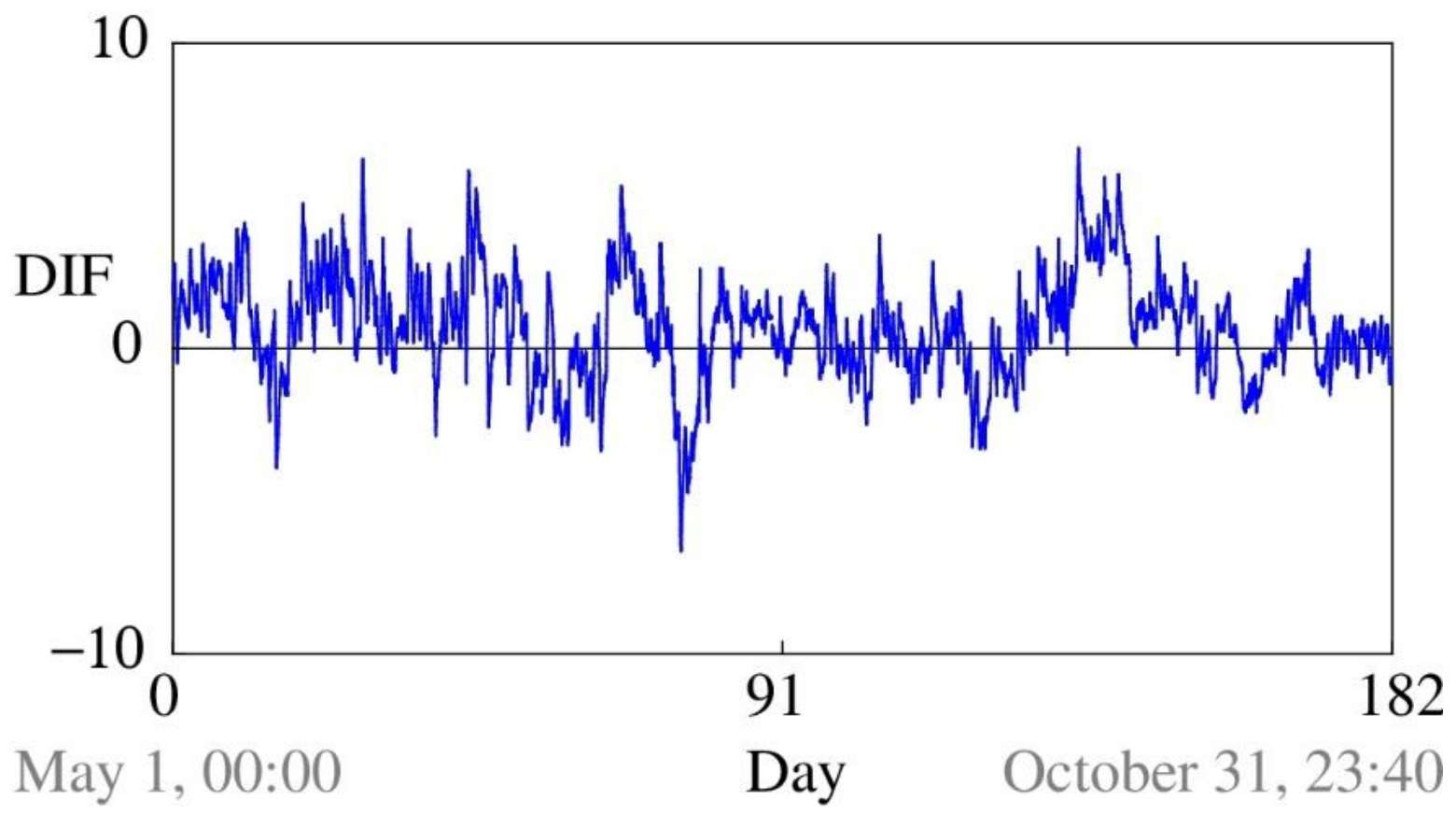

**Figures A3.** Difference (DIF) in water temperatures at Yumura between 2024-2023. The x-axis represents the time in days, while the y-axis represents the temperature difference in degrees Celsius (°C).

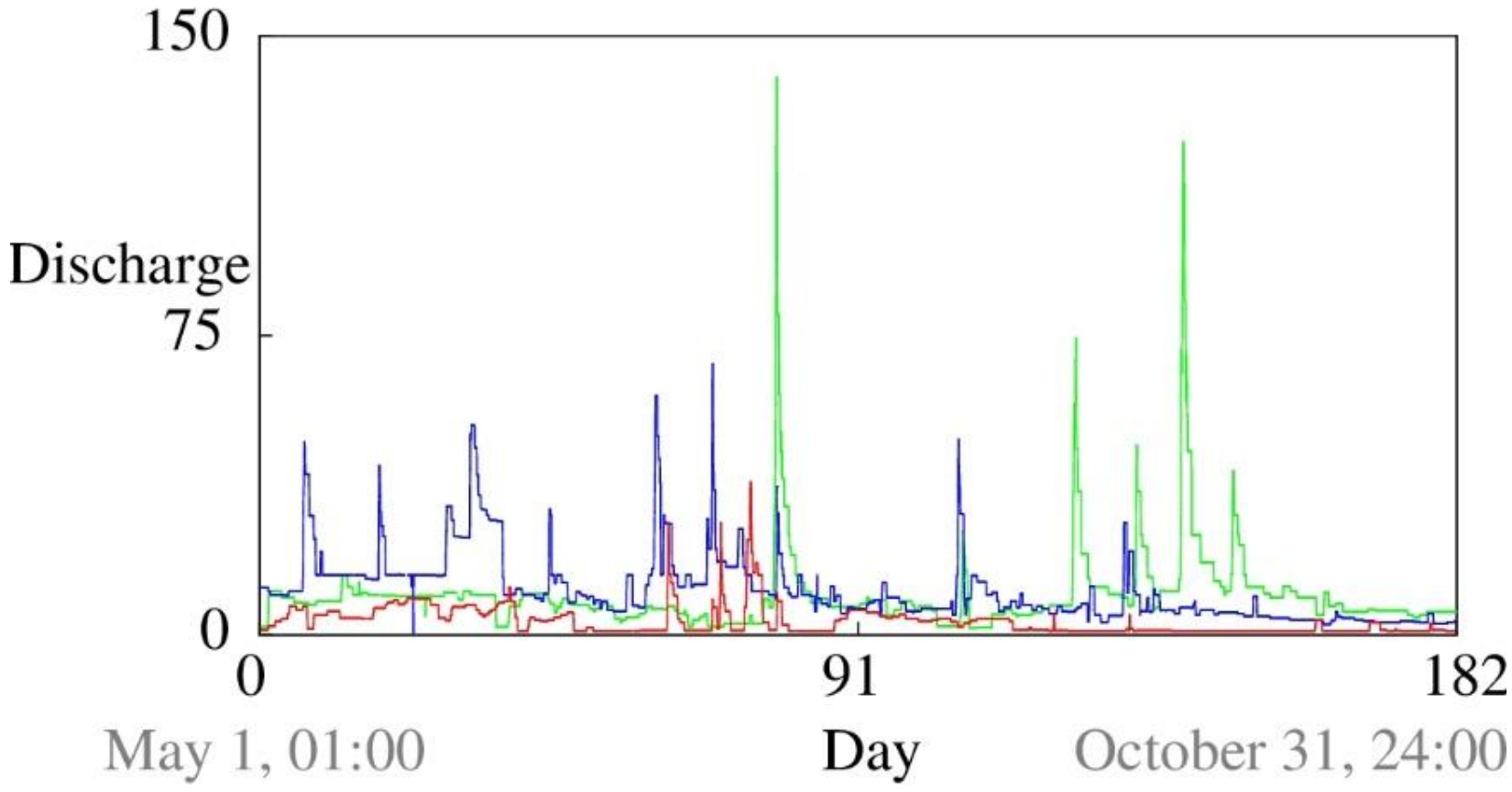

**Figure A4.** Water discharge ($m^3/s$) from the Obara Dam in 2022 (green), 2023 (blue), and 2024 (red). The time unit is day.

### A.3 Identified parameter values for prescribed $\sigma$

**Tables A3-A6** summarize the identified parameter values of $\alpha$, $\beta$, $u$, and $r$ for the Gamma case. **Table A3** presents the parameter $\sigma$ (assumed to be positive without loss of generality) as part of the identification process, while **Tables A4-A6** show the results for scenarios where its value is prescribed at specific values.

**Table A3.** Identified parameter values of the stochastic version of the proposed model: Gamma case.

| | $\alpha$ (-) | $\beta$ (g) | $u$ (-) | $r$ (1/day) | $\sigma$ (1/day$^{1/2}$) |
|---|---|---|---|---|---|
| 2017 | 8.47.E+00 | 5.28.E+01 | 3.22.E-01 | 3.12.E-03 | 5.96E-05 |
| 2018 | 9.59.E+00 | 1.16.E+01 | 2.20.E-01 | 1.42.E-02 | 1.04E-06 |
| 2019 | 9.60.E+00 | 1.37.E+01 | 1.84.E-01 | 1.26.E-02 | 6.77E-07 |
| 2023 | 6.18.E+00 | 1.63.E+01 | 3.11.E-01 | 1.39.E-02 | 7.57E-05 |
| 2024 | 1.01.E+01 | 3.27.E+06 | 1.01.E-02 | 1.17.E-05 | 8.71.E-05 |

**Table A4.** Identified parameter values of the model with stochastic fluctuation: $\sigma = 0.01$ (1/day$^{1/2}$).

| | $\alpha$ (-) | $\beta$ (g) | $u$ (-) | $r$ (1/day) |
|---|---|---|---|---|
| 2017 | 8.47.E+00 | 5.36.E+01 | 3.19.E-01 | 3.03.E-03 |
| 2018 | 9.59.E+00 | 1.16.E+01 | 2.19.E-01 | 1.42.E-02 |
| 2019 | 9.60.E+00 | 1.37.E+01 | 1.83.E-01 | 1.26.E-02 |
| 2023 | 6.18.E+00 | 1.63.E+01 | 3.09.E-01 | 1.39.E-02 |
| 2024 | 1.01.E+01 | 4.71.E+02 | 1.92.E-01 | 1.98.E-04 |

**Table A5.** Identified parameter values of the model with stochastic fluctuation: $\sigma = \sqrt{0.001}$ (1/day$^{1/2}$).

| | $\alpha$ (-) | $\beta$ (g) | $u$ (-) | $r$ (1/day) |
|---|---|---|---|---|
| 2017 | 8.47.E+00 | 6.29.E+01 | 2.94.E-01 | 2.26.E-03 |
| 2018 | 9.59.E+00 | 1.17.E+01 | 2.11.E-01 | 1.38.E-02 |
| 2019 | 9.60.E+00 | 1.38.E+01 | 1.74.E-01 | 1.23.E-02 |
| 2023 | 6.18.E+00 | 1.64.E+01 | 2.99.E-01 | 1.36.E-02 |
| 2024 | 1.01.E+01 | 4.34.E+01 | 4.16.E-01 | 2.12.E-04 |

**Table A6.** Identified parameter values of the model with stochastic fluctuation: $\sigma = 0.1$ (1/day$^{1/2}$).

| | $\alpha$ (-) | $\beta$ (g) | $u$ (-) | $r$ (1/day) |
|---|---|---|---|---|
| 2017 | 8.47.E+00 | 3.11.E+01 | 2.07.E-01 | 2.83.E-03 |
| 2018 | 9.59.E+00 | 1.22.E+01 | 1.31.E-01 | 1.15.E-02 |
| 2019 | 9.60.E+00 | 1.46.E+01 | 9.67.E-02 | 9.75.E-03 |
| 2023 | 6.18.E+00 | 1.67.E+01 | 2.00.E-01 | 1.17.E-02 |
| 2024 | 1.01.E+01 | 8.62.E+00 | 5.79.E-01 | 3.29.E-03 |

**A.4 Orlicz space**

We give a definition of an Orlicz space in the domain $(0,+\infty)$ based on the literature [31] under a minimal setting suited to this study. Assume that a PDF $p$ of a positive random variable is given. For a convex and increasing function $\Phi:(0,+\infty)\rightarrow[0,+\infty)$ such that $\Phi(0)=0$, the Orlicz space $\mathbb{L}_{\Phi}$ associated with $\Phi$ is defined as follows:

$$\mathbb{L}_{\Phi}=\left\{f:(0,+\infty)\rightarrow[0,+\infty);\text{there is some } z>0 \text{ such that } \int_{0}^{+\infty}\Phi\left(f\left(\frac{x}{z}\right)\right)p(x)\mathrm{d}x<+\infty\right\}. \qquad (44)$$

The Orlicz function $\Phi=\exp(\cdot)$ and $f\left(\frac{x}{z}\right)=\frac{x}{z}$ was considered in the main text.